\documentclass{article}
\usepackage{amsmath}
\usepackage{amssymb}
\usepackage{anysize}
\usepackage{appendix}
\usepackage{authblk}
\newcommand{\be}{\begin{equation}}
\newcommand{\ee}{\end{equation}}
\newcommand{\ben}{\begin{equation*}}
\newcommand{\een}{\end{equation*}}
\newcommand{\bea}{\begin{eqnarray}}
\newcommand{\eea}{\end{eqnarray}}
\newcommand{\nn}{\nonumber}
\newcommand{\bean}{\begin{eqnarray*}}
\newcommand{\eean}{\end{eqnarray*}}
\newcommand{\dd}{\mathrm{d}}
\newcommand{\p}{\partial}
\newcommand{\rt}{\right}
\newcommand{\lt}{\left}
\newcommand{\ba}{\begin{array}}
\newcommand{\ea}{\end{array}}

\newcommand{\tg}{\tilde g}
\newcommand{\tx}{\tilde x}
\newcommand{\ta}{\tilde\alpha}
\newcommand{\bvg}{\breve g}
\newcommand{\bva}{\breve\alpha}
\newcommand{\bvA}{\breve A}
\newcommand{\s}{\sigma}
\newcommand{\la}{\Lambda}
\newcommand{\lm}{\lambda}
\newcommand{\kk}{k_{2,0}^2}
\newcommand{\mb}{\mathbf}

\title{Renormalization group analysis of the singularity structure of effective potentials}

\author[1]{T.~Hanif}
\author[2]{N.~Ishtiaque}
\affil[1] {Department of Theoretical Physics, University of Dhaka, Dhaka-1000, Bangladesh}
\affil[2] {Department of Physics, University of Dhaka, Dhaka-1000, Bangladesh}
\begin{document}
\maketitle
	\begin{abstract}
		Using the renormalization group techniques it was previously shown that the perturbative effective potential in the $\mathcal{O}(N)$ symmetric $\phi^4$ theory, massless scalar electrodynamics as well as in the conformal limit of the standard model can be uniquely determined in terms of the known MS scheme renormalization group functions. Furthermore, re-summation of the leading order corrections plus portions of these higher order contributions to the effective potential in the $\mathcal{O}(N)$ symmetric $\phi^4$ theory shows a peculiar shift in the usual "Landau singularity" apparent in the individual leading order corrections. In this work, we have further
		investigated this shift by extending the result for theories having multiple couplings. We have shown that the singularity structure of the effective potential as seen by summing up portions of the $V_{N^{P+n}LL}$ for any finite $P$ is altered as expected when these perturbative contributions are summed to all orders but more significantly we found that this shift in the singularity is completely determined by the one and two loop beta functions only. We argue that this is a general result which applies to the conformally invariant standard model.
	\end{abstract}
	
	\section*{Keywords}
renormalization group equations; renormalization group functions; Coleman-Weinberg renormalization scheme; effective potential; standard model; Landau singularity.

	%\maketitle

\section{INTRODUCTION}
	It was shown that the perturbative effective potential for the $\mathcal{O}(N)$ symmetric $\phi^4$ theory and massless
	scalar electrodynamics can be determined uniquely to any loop order using the renormalization group equations in 
	conjunction with the Coleman-Weinberg renormalization condition, provided that the MS scheme renormalization group 
	functions are known to that order \cite{unique}. In other words the essential inputs for this calculation are the known
	MS scheme RG functions. But the renormalization group equations constructed from the ansatz for the effective potential
	in the Coleman-Weinberg scheme involves RG functions in that scheme and necessitates scheme conversion for which a 
	systematic methodology was also developed in \cite{unique}. For example, in case of the $\mathcal{O}(N)$ symmetric
	$\phi^4$ theory if one uses the RG equation in conjunction with the CW renormalization condition (which provides the
	necessary boundary conditions) one can iteratively solve the coupled differential equations for the $N^PLL$ order
	contributions to $V$. The effective potential thus obtained is called RG improved because it contains more terms (the
	leading logarithmic terms) than the order corresponding to the required RG functions necessary to obtain $V_{N^PLL}$.
	For example, the $(n+1)$-loop effective potential fixes the $(n+1)$-loop beta function and this in turn can be used to
	exactly sum portions of the effective potential beyond it without the appearance of any unknown parameter \cite{unique,
	higgs, summing, improv, eff_pot_phi4}. This amounts to summing the $n$ times next to the highest power of
	$\ln\frac{\phi^2}{\mu^2}$ at each order of the perturbation theory and we denote it as $V_{N^nLL}$. It is necessary to know
	the $(P+1)$-loop RG functions to determine $V_{N^PLL}$ and the $(P+1)$-loop order effective potential can be extracted 
	from the sum $V_{LL} + V_{NLL} + \cdots + V_{N^PLL}$ for consistency check. For example, the two-loop $V$ fixes the
	two-loop beta function which in turn fixes the next-to leading log corrections to $V$ using the renormalization group
	equations in conjunction with the CW renormalization condition. This two-loop $V$ can be completely recovered from the
	sum $V_{LL} + V_{NLL}$ and fixing the logarithm-independent term using the CW condition. There are other ways to calculate
	the $n$-loop beta functions and thus the RG equations might have been used to calculate the $n$-loop $V$ ab-initio
	\cite{unique}. Once we know $V_{LL}$ and $V_{NLL}$, we can iteratively solve the RG equations to determine $V_{N^2LL}$,
	provided that we know the three-loop beta function and so on. The method was further extended for mass less scalar
	electrodynamics and the Standard Model having conformal invariance \cite{unique, higgs}.

	It was shown \cite{unique} that all the individual leading logarithmic terms in the $\mathcal{O}(N)$ symmetric $\phi^4$ 
	theory has the usual "Landau singularity" at $1-\frac{b_2}{2} \lambda \ln\frac{\phi^2}{\mu^2} = 0$ where $b_2$ is the
	coefficient of the one-loop CW or MS scheme beta function. But it was also demonstrated that summing portions of $V$ that
	consists of terms with highest powers of logarithms from all orders shifts this singularity at $w = 0$. Obviously, this
	shift of singularity observed in \cite{unique} can neither be its unique nor special feature and should in principle 
	hold as a limiting case of massless scalar electrodynamics (MSED) as $\alpha \rightarrow 0$. 

	In the present work this result of $V$ having a peculiar singularity structure is extended for the simplest case where we
	have more than just the quartic scalar coupling or MSED by finding a generic "Landau singularity"
	at $w(\alpha,g) = 0$. Furthermore, we also have shown that this singularity is shifted uniquely by an amount that is completely determined by just the one  and two-loop beta functions for both the couplings.  This is the first step towards a generalization to the Standard Model in the conformal limit for
	which the expansion of $V$ is a straightforward extension of MSED \cite{higgs}. We argued that this is in principle possible
	and mentioned the key points required for this generalization. It would be however, certainly very interesting and useful to
	find a meticulously thorough and rigorous proof.

	It was also shown \cite{summing, improv, eff_pot_phi4} that an alternate summation or rearrangement of the contributions to
	$V$ in powers of $\ln\frac{\phi^2}{\mu^2}$ with coefficients being dependent solely on the coupling lamba with contributions
	coming from all orders of the loop expansion is possible. The log-independent piece of this expansion can be fixed by the
	condition $\lt.\frac{\dd V}{\dd \phi}\rt|_{\phi = \nu} = 0$ and, combined with the RG equations this makes $V$ independent of 
	$\phi$, unless $\nu = 0$ (in which case obviously there is no spontaneous symmetry breaking). This argument for the effective
	potential being flat was given in the context of the $\overline{\mbox{MS}}$ scheme in \cite{improv, eff_pot_phi4} and CW scheme in 
	\cite{summing}. This result was supported by showing the individual leading order contributions to $V$ for the $\mathcal{O}(N)$
	symmetric $\phi^4$ theory becoming progressively less dependent on $\phi$, both in the CW and $\overline{\mbox{MS}}$ scheme
	\cite{summing} and also from a simplified pure $\mathcal{O}(4)$ scalar field theory obtained from the Standard Model by setting all
	the couplings except $\lambda$ to zero \cite{higgs}. Again, in the present work this remarkable result is generalized for MSED in 
	the CW scheme.
	
	This article is organized as follows. In section $2$ we start by giving a brief overview of the simplest conformally invariant gauge theory beyond a single coupling containing a single real scalar field or massless scalar electrodynamics\cite{unique}. Renormalization scale independence of the perturbative effective potential in the Coleman-Weinberg scheme gives rise to the RG equations. Following that, in section $3$, we find explicit 
	power series forms of $V_{LL}$ and $V_{NLL}$ (of MSED) using the method of characteristics which closely follows \cite{unique, higgs}. 
	To find the power series form we introduce some new notations which facilitate the manipulation of the power series involved. In section
	$4$ we make an ansatz about the power series form of $V_{N^PLL}$ as a generalization of our results for $V_{LL}$
	and $V_{NLL}$. The generalized form of $V_{N^PLL}$ reduces to the known expression of $V_{N^PLL}$ for the scalar $\phi^4$ theory in 
	the limit $\alpha \to 0$. This power series form of $V_{N^PLL}$ gives us a power series form of the effective potential $V$. We then
	substitute this form of $V$ in the RG equation which gives us a recurrence relation for the coefficients of the power series. In
	section $5$ we make a partition of the infinite power series of $V$ in a suitable way and deduce some
	asymptotic characteristics of the coefficients which allows us to determine the radius of convergence of the series, which gives us
	the singularity of $V$. In section $6$ we outline some key points in generalizing our results to a theory with multiple
	couplings. Finally in section $7$ we show that an alternate summation of the power series of the effective potential leads to
	the result that the RG improved effective potential of MSED is independent of the scalar field which is consistent with the potential becoming flatter with added contributions from leading logarithm terms \cite{higgs}.
	%\clearpage
	
	%\clearpage

\section{A BRIEF OVERVIEW}
	We start by considering the Lagrangian of massless scalar electrodynamics:
	\be
		\mathcal{L} = \frac{1}{2} (\p_{\mu} + ieA_{\mu})\phi^*(\p^{\mu} - ieA^{\mu})\phi - \frac{1}{4}(\p_{\mu}A_{\nu}-\p_{\nu}A_{\mu})(\p^{\mu}A^{\nu}-\p^{\nu}A^{\mu}) - \frac{ g}{4!}(\phi\phi^*)^2 \nn
	\ee
	where, a complex scalar $\phi$ is coupled to a $U(1)$ gauge field $A_\mu$ with
coupling $e$.

	The effective action for this theory can be expanded as follows:
	\be
		\Gamma = \int d^4x \left[ -V(\phi) + \frac{1}{2} Z(\phi)\left|(\partial_\mu - ieA_\mu)\phi\right|^2 -  \frac{1}{4} H(\phi)(\partial_\mu A_\nu - \partial_\nu A_\mu)^2 + \ldots\right].
	\ee
	
	Now, the coupling $g$ is renormalized in such a way that the effective potential $V$ satisfies the CW renormalization condition:
	\be \lt.\frac{\dd^4 V}{\dd \phi^4}\rt|_{\phi=\mu} = g \label{e5} \ee 
	supplemented by the conditions
	\bea
	H(\phi = \mu) = 1 = Z(\phi = \mu).
	\eea
	
	In the Coleman-Weinberg renormalization scheme the ansatz for the perturbative  effective potential has the following form\cite{unique, higgs}:
	\bea
		V(g,\alpha,\phi,\mu) &=& \sum_{n=1}^{\infty} \sum_{k=0}^{n-1} \sum_{r=0}^n T_{n-r,r,k} g^{n-r} \alpha^r L^k \phi^4;\quad [L = \mathrm{ln}(\phi^2/\mu^2), \alpha = e^2] \label{e40}\\
		&=& \sum_{n=1}^{\infty} \sum_{k=0}^{n-1} P_n^k( g,\alpha) L^k \phi^4 \label{e4}
	\eea
	Defining, \be V_{N^PLL} = \sum_{k=0}^{\infty} P_{k+p+1}^k L^k \phi^4 \nn\ee  we can write,
	\be V = \sum_{P=0}^\infty V_{N^PLL} \nn\ee

	The independence of the effective potential $V$ on the renormalization scale parameter $\mu$ leads to the renormalization group  equation:
	\bea
		\mu \frac{\dd V}{\dd \mu} &=& 0 \nn\\
		\Rightarrow \lt(\mu\frac{\p}{\p\mu} + \beta^g\frac{\p}{\p g} + \beta^{\alpha}\frac{\p}{\p\alpha} + \gamma\phi\frac{\p}{\p\phi}\rt) V &=& 0 \label{e6}
	\eea
	where,
	\begin{subequations}\bea
		\beta^g( g,\alpha) = \mu\frac{\dd g}{\dd \mu} = \sum_{n=2}^{\infty} \beta_n^g &,& \beta_n^g(\alpha, g) = \sum_{r=0}^n b_{n-r,r}^g g^{n-r}\alpha^r \\
		\beta^\alpha( g,\alpha) = \mu\frac{\dd \alpha}{\dd \mu} = \sum_{n=2}^{\infty} \beta_n^\alpha &,& \beta_n^\alpha(\alpha, g) = \sum_{r=0}^n b_{n-r,r}^\alpha g^{n-r}\alpha^r\\
		\gamma( g,\alpha) = \frac{\mu}{\phi}\frac{\dd \phi}{\dd \mu} = \sum_{n=1}^{\infty} \gamma_n &,& \gamma_n(\alpha, g) = \sum_{r=0}^n \tau_{n-r,r} g^{n-r}\alpha^r
	\eea\label{e7}\end{subequations}
	
	%In the following section we find explicit power series forms of $V_{LL}$ and $V_{NLL}$ (of MSED) using the method of characteristics which closely follows \cite{unique, higgs}. Next, we make an ansatz about the power series form of $V_{N^PLL}$ as a generalization of our results for $V_{LL}$ and $V_{NLL}$. The generalized form of $V_{N^PLL}$ reduces to the known expression of $V_{N^PLL}$ for the scalar $\phi^4$ theory in the limit $\alpha \to 0$. This power series form of $V_{N^PLL}$ gives us a power series form of the effective potential $V$. We then substitute this form of $V$ in the RG equation which gives us a recurrence relation for the coefficients of the power series. Subsequently we make a partition of the infinite power series of $V$ in a suitable way and deduce some asymptotic characteristics of the coefficients which allows us to determine the radius of convergence of the series, which in turn give us the singularity of $V$.

\section{EVALUATION OF $V_{LL}$ AND $V_{NLL}$}
	The method of characteristics can be used to find the expressions of $V_{LL}$ and $V_{NLL}$ \cite{unique, higgs}.
	 We first define the function,
		\be w_{n+k}^k(\tg(t), \ta(t), t) = \mathrm{exp}\lt[ 4\int_0^t \gamma_1(\tg(\tau), \ta(\tau), \tau) \mathrm{d}\tau \rt] P_{n+k}^k(\tg(t), \ta(t)) \label{e14}\ee
		where $\ta(t)$ and $\tg(t)$ are defined as the solutions of the following differential equations:
		\begin{subequations} \label{e15}\bea
			\frac{\dd \ta(t)}{\dd t} = \beta_2^{\alpha}(\tg(t),\ta(t)) &,& (\ta(0)=\alpha) \\
			\frac{\dd \tg(t)}{\dd t} = \beta_2^{ g}(\tg(t),\ta(t)) &,& (\tg(0)= g)
		\eea\end{subequations}
		From (\ref{e14}) and (\ref{e15}) it follows that,
		\be \frac{\dd}{\dd t} w_{n+k}^k(\tg, \ta, t) = \lt(\beta_2^g(\tg, \ta)\frac{\p}{\p \tg} + \beta_2^{\alpha}(\tg, \ta)\frac{\p}{\p \ta} + 4\gamma_1(\tg, \ta) \rt) w_{n+k}^k(\tg, \ta, t) \label{e16} \ee

		Now we look for recurrence relations for $w$. Substituting
		(\ref{e4}) in (\ref{e5}) we get,
		\be 24P_n^0 + 100P_n^1 + 280P_n^2 + 480P_n^3 + 384P_n^4 = g \label{e8}\ee
		For $n=1$ we find, \be P_1^0 = \frac{g}{24} \label{e9} \ee
		Substituting (\ref{e4}) and (\ref{e7}) in (\ref{e6}) we get,
		\be \sum_{n=1}^{\infty} \sum_{k=0}^{n-1} \lt[-2kP_n^kL^{k-1} + \sum_{m=2}^{\infty} \lt(\beta_m^g \frac{\p P_n^k}{\p g} + \beta_m^{\alpha} \frac{\p P_n^k}{\p \alpha}\rt) L^k + \sum_{m=1}^{\infty} (4\gamma_m P_n^k L^k + 2k\gamma_m P_n^k L^{k-1}) \rt] \phi^4 = 0 \label{e10} \ee
		At order $n+2$ in couplings and order $n$ in $L$ we find,
		\be P_{n+2}^{n+1} = \frac{1}{2(n+1)} \lt( \beta_2^g\frac{\p}{\p g} + \beta_2^{\alpha}\frac{\p}{\p \alpha} + 4\gamma_1 \rt) P_{n+1}^n \label{e11} \ee
		$P_2^1$ can be calculated from (\ref{e11}) using $P_1^0$. Then putting $n=2$ in (\ref{e8}) we get, \be P_2^0 = -\frac{25}{6} P_2^1 \label{e12} \ee
		Now from (\ref{e10}), at order $n+3$ in couplings and order $n$ in $L$,
		\be P_{n+2}^n - \gamma_1P_{n+1}^n = \frac{1}{2n} \lt[\lt( \beta_2^g\frac{\p}{\p g} + \beta_2^{\alpha}\frac{\p}{\p \alpha} + 4\gamma_1 \rt) P_{n+1}^{n-1} + \lt( \beta_3^g\frac{\p}{\p g} + \beta_3^{\alpha}\frac{\p}{\p \alpha} + 4\gamma_2 \rt) P_{n-1}^n \rt] \label{e13} \ee
		This fixes all the contributions to $V_{LL}$ and $V_{NLL}$ in terms of $\beta_2^g, \beta_3^g, \beta_2^{\alpha}, \beta_3^{\alpha}, \gamma_1$ and $\gamma_2$. This procedure can be continued to find recurrence relations beyond equations (\ref{e11}, \ref{e13}).
		
		 From (\ref{e11}) and (\ref{e14}) it follows,
		\be w_{n+1}^n(\tg, \ta, t) = \frac{1}{2n} \lt( \beta_2^g\frac{\p}{\p g} + \beta_2^{\alpha}\frac{\p}{\p \alpha} + 4\gamma_1 \rt) w_n^{n-1}(\tg, \ta, t) \label{e17} \ee
		
		We now define,
		\be V_{N^PLL}(t) = \sum_{n=0}^{\infty} w_{n+p+1}^n(\tg(t), \ta(t), t) \tilde L^n \phi^4 \label{e33}\ee
		Using this definition and the recurrence relations for $w$ obtained in the manner outlined earlier, functional forms
		for $V_{N^PLL}$ can be found. In particular, we evaluate the forms of $V_{LL}$ and $V_{NLL}$ in the following 
		subsections. But before that, we find the explicit forms of the characteristic functions $\ta$ and $\tg$ since they
		will appear in the subsequent calculations of the effective potential.
		\vspace{0.6cm}
		
		%\noindent\textbf
		\subsection	{Explicit forms of the characteristic functions}
		\vspace{0.2cm}
		
			\noindent The beta functions of the model upto 1-loop are :
			\begin{subequations} \label{e1} \bea
				\beta^{ g} &=& \frac{1}{16\pi^2} (5 g^2 - 12 g\alpha + 24\alpha^2) + \cdots \\
				\beta^{\alpha} &=& \frac{\alpha^2}{24\pi^2} + \cdots
			\eea\end{subequations}
			From (\ref{e15}) and (\ref{e1}),
			\begin{subequations} \bea
				\frac{\dd \ta}{\dd t} &=& \frac{\ta^2}{24\pi^2} \label{e2a}\\
				\frac{\dd \tg}{\dd t} &=& \frac{1}{16\pi^2} (5\tg^2 - 12\tg\ta + 24\ta^2) \label{e2b}
			\eea \end{subequations}
			Solving (\ref{e2a}), \be \ta(t) = \frac{\alpha}{1-\frac{\alpha t}{24\pi^2}} \label{alpha_t}\ee
			Solving (\ref{e2b}),
			\be \tg(t) = \ta(t)\frac{\sqrt{719}g\, \mathrm{cos}\lt(\frac{\sqrt{719}}{2}\mathrm{ln}\frac{\ta(t)}{\alpha}\rt) + (72\alpha-19g)\, \mathrm{sin}\lt(\frac{\sqrt{719}}{2}\mathrm{ln}\frac{\ta(t)}{\alpha}\rt)}{\sqrt{719}\alpha\, \mathrm{cos}\lt(\frac{\sqrt{719}}{2}\mathrm{ln}\frac{\ta(t)}{\alpha}\rt) + (19\alpha-15g)\, \mathrm{sin}\lt(\frac{\sqrt{719}}{2}\mathrm{ln}\frac{\ta(t)}{\alpha}\rt)} \ee
			$\tg(t)$ has a simple pole at,
			\be
				t = \frac{24\pi^2}{\alpha}\lt(1 - \mathrm{exp}\lt[\frac{-2}{\sqrt{719}} \mathrm{tan}^{-1} \frac{\sqrt{719}\alpha}{15g - 19\alpha}\rt]\rt) := w(\alpha,g)\nn
			\ee
			Expanding $\ta(t)$ and $\tg(t)$ around $w$ we can write:
			\bea
				\ta(t) &=& C^{\alpha}_0 \ta(w) + C^{\alpha}_1 \ta(w)^2 (t-w) + \mathcal{O}\lt(\ta(w)^3 (t-w)^2\rt) \label{e65}\\
				\tg(t) &=& \frac{C^g_{-1}}{t-w} + C^g_0 \ta(w) + C^g_1 \ta(w)^2 (t-w) + \mathcal{O}\lt(\ta(w)^3 (t-w)^2\rt) \label{e66}
			\eea
			where,
			\be \lim_{\alpha \to 0} \,\frac{C^g_{-1}}{t-w(\alpha,g)} = \frac{g}{1-\frac{5gt}{16\pi^2}} \nn\ee
			We note that,
			\bea
				\lim_{\alpha \to 0} w(\alpha, g) &=& \frac{16\pi^2}{5g} \label{limit_w}\\
				\mbox{and,}\qquad \lim_{\alpha \to 0} \tg(t) &=& \frac{g}{1-\frac{5gt}{16\pi^2}},\label{limit_g}
			\eea
			which is the correct single coupling case. We notice that equations (\ref{e65}) and (\ref{e66}) are power series
			in $\ta(w)$ and $(t-w)$ where the $n$-th term of the series is proportional to $\ta(w)^{n+m+k}(t-w)^{n+k}$ for 
			some $m, k$ that are fixed for a series. Since the effective potential will depend on these characteristic 
			functions, the potential will also be expressible as such a series and therefore we introduce some new notations to
			write these type of series concisely and to manipulate them more easily.
			\vspace{0.6cm}
		
	\subsection{New notations}
			If $S = \{s_1,s_2,\cdots\}$ then define,
			\bea L[S] &:=& \mbox{Any linear combination of the elements } s_1, s_2,\cdots\nn\\
				&\equiv& \sum_{s_i \in S} c_is_i\quad \mbox{for any\, } c_i \in \mathbb{R} \nn\\
				\,[m]_k^t &:=& L[\{\ta(w)^{i+m} (t-w)^i\; |\; k \leq i < \infty\}] \nn\\
				&\equiv& \sum_{i=k}^\infty c_i \ta(w)^{i+m} (t-w)^i \label{e95}
			\eea
			$[m]_k^t$ is a type of generic series in the variable $t$. $m$ and $k$ serves as parameters of the series and the
			series depends on the couplings through the dependence of $\omega$ on the couplings and the $\alpha$ in $\ta$. 
			Properties of the entities $[m]_k^t$ such as addition, multiplication, inverse and integrations are discussed in
			Appendix \ref{app1}. 
			
			We rewrite the characteristic functions, their derivatives and the beta functions using these notations,
			\be\begin{split} \tg(t) &= [1]_{-1}^t \\
				\ta(t) &= [1]_0^t
			\end{split}\label{e67}\ee
			\be\begin{split} \beta_n^{\alpha} (\ta, \tg) &= L[\{\ta^{n-i}\tg^i\; |\; 0 \leq i \leq n-1\}] = [n]_0 + [n]_{-1} + \cdots + [n]_{-n+1} = [n]_{-n+1} \\
				\beta_n^{g} (\ta, \tg) &= L[\{\ta^{n-i}\tg^i\; |\; 0 \leq i \leq n\}] = [n]_0 + [n]_{-1} + \cdots + [n]_{-n} = [n]_{-n} \\
				\gamma_1 (\ta, \tg) &\propto \ta = [1]_0
			\end{split}\label{e69}\ee
			With the necessary tools in hand we now move on to the the evaluation of $V_{LL}$ and $V_{NLL}$.

	\subsection{Evaluation of $V_{LL}$} \label{secVLL}
		Putting $p=0$ in equation (\ref{e33}),
		\be V_{LL}(t) = \sum_{n=0}^{\infty} w_{n+1}^n(\tg(t), \ta(t), t) \tilde L^n \phi^4 \label{e18}\ee
		where $$\tilde L = \mathrm{log}\frac{\phi^2}{\tilde \mu(t)^2}$$ with $$\frac{\dd \tilde\mu(t)}{\dd t} = \tilde\mu(t);\quad \tilde\mu(0) = \mu$$
		From (\ref{e16}) and (\ref{e17}),
		\be w_{n+1}^n(\tg, \ta, t) = \frac{1}{2^nn!} \frac{\dd^n}{\dd t^n} w_1^0(\tg, \ta, t) \label{e19}\ee
		Therefore, (\ref{e18}) becomes,
		\be
			V_{LL}(t) = \sum_{n=0}^{\infty} \frac{\tilde L^n}{2^nn!} \frac{\dd^n}{\dd t^n} w_1^0(\tg, \ta, t) \phi^4 = w_1^0\lt(\tg\lt(t+\frac{\tilde L}{2}\rt), \ta\lt(t+\frac{\tilde L}{2}\rt), t+\frac{\tilde L}{2} \rt) \phi^4 \nn
		\ee
		Define, $$l := \frac{L}{2}.$$
		Now, \bea
			 V_{LL} = V_{LL}(t=0) &=& w_1^0\lt(\tg\lt(l\rt), \ta\lt(l\rt), l \rt) \phi^4 \nn\\
			&=& \mathrm{exp}\lt[ 4\int_0^l \gamma_1(\tg(\tau), \ta(\tau), \tau) \mathrm{d}\tau \rt] P_1^0\lt(\tg\lt(l\rt), \ta\lt(l\rt)\rt) \phi^4 \nn\\
			&=& \mathrm{exp}\lt[ 4\int_0^l \gamma_1(\tg(\tau), \ta(\tau), \tau) \mathrm{d}\tau \rt] \frac{\tg\lt(l\rt)}{24} \phi^4 \nn\\
			&=& K \frac{\tg\lt(l\rt)}{24} \phi^4;\quad \lt[K := \mathrm{exp}\lt[ 4\int_0^l \gamma_1(\tg(\tau), \ta(\tau), \tau) \mathrm{d}\tau \rt] = (1 - b_{0,2}^\alpha \alpha l)^{-\frac{4 \tau_{0,1}}{b_{0,2}^\alpha}} \rt] \nn
		\eea
		Using (\ref{e67}) we also note that,
		\be K^{-1}V_{LL} = [1]_{-1}^l \phi^4 \label{e71}\ee

	\subsection{Evaluation of $V_{NLL}$} \label{secVNLL}
		For notational convenience let,
		\bea \tx_1(t) &:=& \ta(t) \nn\\
			\tx_2(t) &:=& \tg(t) \nn
		\eea
		From (\ref{e13}), (\ref{e14}), (\ref{e16}) and (\ref{e19}),
		\be w_{n+2}^n = \frac{1}{2n} \lt[\frac{\dd}{\dd t} w_{n+1}^{n-1} + D(t) w_n^{n-1} \rt] \label{e20}\ee
		where \be D(t) = -\gamma_1 \lt( \beta_2^{x_1} \frac{\p}{\p\tx_1} + \beta_2^{x_2} \frac{\p}{\p\tx_2} - 4\gamma_1 \rt) + \lt( \beta_3^{x_1} \frac{\p}{\p\tx_1} + \beta_3^{x_2} \frac{\p}{\p\tx_2} - 4\gamma_2 \rt). \ee
		Iterating (\ref{e20}),
		\be w_{n+2}^n = \frac{1}{2^nn!} \lt[ \frac{d^n}{dt^n} w_2^0 + \lt( \frac{d^{n-1}}{dt^{n-1}} D(t) + \frac{d^{n-2}}{dt^{n-2}} D(t) \frac{d}{dt} + \cdots + D(t) \frac{d^{n-1}}{dt^{n-1}}\rt) w_1^0 \rt] \label{e34}\ee
		One can inductively prove the identity
		\be \lt( \frac{\dd^{n-1}}{\dd t^{n-1}} f + \frac{\dd^{n-2}}{\dd t^{n-2}} f \frac{\dd}{\dd t} + \ldots + \frac{\dd}{\dd t} f \frac{\dd^{n-2}}{\dd t^{n-2}} + f \frac{\dd^{n-1}}{\dd t^{n-1}}\rt)g = \frac{\dd^n}{\dd t^n} (\phi g) - \phi \frac{\dd^n}{\dd t^n} g \qquad (\ \frac{\dd\phi}{\dd t} := f \ ). \label{e30}\ee
	
		Now lets change the characteristic functions from $\tx_1, \tx_2$ to $y_1, y_2$ such that,
		\be\begin{split}
			y_1(\tx_1, \tx_2) &= \tx_1 \\
			\mbox{and } y_2(\tx_1, \tx_2) \mbox{ satisfies,}\quad \frac{\dd y_2}{\dd t} &= \beta_2^{x_1} \frac{\p y_2}{\p\tx_1} + \beta_2^{x_2} \frac{\p y_2}{\p\tx_2} = \frac{y_2^2}{wg} \\
			\mbox{with, }\quad y_2(x_1, x_2) &= \frac{16\pi^2}{5w};
		\end{split} \label{e21}\ee
		where, $x_1 := \tx_1(0) = \alpha,\, x_2 := \tx_2(0) = g$. The benefit of these characteristic functions is
		that unlike (\ref{e2b}), (\ref{e21}) is decoupled.
		
		Solving the ODE
		\be \frac{\dd}{\dd t} y_2(\tx_1(t),\tx_2(t)) = \frac{5y_2(\tx_1(t),\tx_2(t))^2}{16\pi^2} \nn\ee
		with the boundary condition $y_2(\tx_1(0),\tx_2(0)) = \frac{16\pi^2}{5w}$ we get,
		\be y_2(t) = \frac{16\pi^2}{5w}\frac{1}{1-tw^{-1}};\qquad [y_2(t) := y_2(\tx_1(t),\tx_2(t))] \nn\ee
		and, \be \lim_{\alpha \to 0} y_2(t) = \frac{g}{1-\frac{5gt}{16\pi^2}} = \lim_{\alpha \to 0} \tg(t);\qquad \lt[\lim_{\alpha \to 0} w(\alpha, g) = \frac{16\pi^2}{5g}\rt] \nn\ee which is the correct single coupling case. And obviously, $\lim_{\alpha \to 0} y_1 = 0$. 
		This ensures that,  what follows is consistent with the single coupling theory to which it must reduce in the limit
		$\alpha \to 0$.
	
		The new characteristic functions satisfy, using (\ref{e21}),
		\be\begin{split}
			\frac{\dd y_1}{\dd t} &= \frac{\dd \tx_1}{\dd t} = b_{0,2}^{x_1} \tx_1^2 = b_{0,2}^{x_1} y_1^2 := \beta_2^{y_1} \\
			\frac{\dd y_2}{\dd t} &= \frac{y_2^2}{w x_2} := b_{2,0}^{y_2} y_2^2 := \beta_2^{y_2}
		\end{split} \label{e22}\ee
		with solutions,
		\be\begin{split}
			y_1(t) &= \frac{y_1(0)}{\displaystyle 1-b_{0,2}^{x_1} y_1(0)\, t} \\
			y_2(t) &= \frac{y_2(0)}{\displaystyle 1-b_{2,0}^{y_2} y_2(0)\, t}
		\end{split}\label{e28}\ee
		Now,
		\bea
			D(t) &=& (-\gamma_1 \beta_2^{x_1} + \beta_3^{x_1}) \frac{\p}{\p \tx_1} + (-\gamma_1 \beta_2^{x_2} + \beta_3^{x_2}) \frac{\p}{\p \tx_2} + 4(\gamma_1^2-\gamma_2) \nn\\
			&=& (-\gamma_1 \beta_2^{x_1} + \beta_3^{x_1}) \frac{\p}{\p y_1} + \lt[ (-\gamma_1 \beta_2^{x_2} + \beta_3^{x_2}) \frac{\p y_2}{\p \tx_2} + (-\gamma_1 \beta_2^{x_1} + \beta_3^{x_1}) \frac{\p y_2}{\p \tx_1} \rt] \frac{\p}{\p y_2} + B(\tx(t)) \nn\\
			&:=& A_i(\tx(t)) \frac{\p}{\p y_i} + B(\tx(t)) \label{e29}
		\eea
		Furthermore, using (\ref{e22}),
		\be \frac{\dd}{\dd t} = \beta_2^{y_i}(y) \frac{\p}{\p y_i} + \frac{\p}{\p t} := \eta_i(y) \frac{\p}{\p y_i} + \frac{\p}{\p t} \nn\ee
		We now note that,
		\bea
			A_i \frac{\p}{\p y_i} \frac{df}{dt} &=& A_i \frac{\p}{\p y_i} \lt(\eta_j \frac{\p}{\p y_j} + \frac{\p}{\p t}\rt) f \nn \\
			&=& A_i \lt[ \lt( \eta_j \frac{\p}{\p y_j} + \frac{\p}{\p t}\rt) \frac{\p f}{\p y_i} + \frac{\p \eta_j}{\p y_i}\frac{\p f}{\p y_j}\rt]\nn \\
			&=& A_i \lt[ \frac{d}{dt} \delta_{ij} + (\mathbf{M})_{ij} \rt]\frac{\p f}{\p y_j} \nn
		\eea
		where \be (\mathbf{M})_{ij} = \frac{\p\eta_j}{\p y_i} = 2 \lt(\begin{array}{cc} b_{0,2}^{x_1}y_1 & 0 \\ 0 & b_{2,0}^{y_2}y_2 \end{array}\rt), \label{e23}\ee
		and hence,  by iterating we obtain
		\be A_i \frac{\p}{\p y_i} \lt(\frac{d}{dt}\rt)^pf = A_i \lt[ \lt( \frac{d}{dt} + \mathbf{M}\rt)^p\rt]_{ij} \frac{\p f}{\p y_j}. \label{e24}\ee
		We now define
		\bea
			(\mathbf{U}(t,0))_{ij} &=& \delta_{ij} + \sum_{n=1}^\infty \int_0^t \dd\tau_1 \int_0^{\tau_1} \dd\tau_2 \ldots \int_0^{\tau_{n-1}}\dd\tau_{n} \lt[ \mathbf{M}(\tau_n)\mathbf{M}(\tau_{n-1}) \cdots \mathbf{M}(\tau_2) \mathbf{M}(\tau_1) \rt]_{ij} \nn\\
			&=& \lt( \begin{array}{cc} \lt(1-b_{0,2}^{x_1}\, y_1(0)\,t\rt)^{-2} & 0 \\ 0 & \lt(1-b_{2,0}^{y_2}\, y_2(0)\,t\rt)^{-2} \end{array} \rt) \label{e32}
		\eea
		It is evident that
		\be \frac{d}{dt}(\mathbf{U} (t,0)f) = \mathbf{U} (t,0) \lt(\frac{d}{dt} + \mathbf{M}\rt) f \label{e26}\ee
		and, 
		\be
			\mathbf{U}^{-1}(t,0) = \lt( \begin{array}{cc} \lt(1-b_{0,2}^{x_1}\, y_1(0)\,t\rt)^2 & 0 \\ 0 & \lt(1-b_{2,0}^{y_2}\, y_2(0)\,t\rt)^2 \end{array} \rt)
		\label{e27}\ee
		Together, eqs. (\ref{e23}-\ref{e27}) show that
		\be A_i \frac{\p}{\p y_i} \lt(\frac{d}{dt}\rt)^p f = A_i \lt[ \mathbf{U}(0,t)\lt(\frac{d}{dt}\rt)^p \mathbf{U}(t,0)\rt]_{ij}\frac{\p}{\p y_j}f. \label{e31}\ee
		We now find that by eqs. (\ref{e30}, \ref{e29}, \ref{e31})
		\bea
			&& \lt(\frac{d^{n-1}}{dt^{n-1}} D(t) + \frac{d^{n-2}}{dt^{n-2}} D(t) \frac{d}{dt} + \ldots + \frac{d}{dt} D(t) \frac{d^{n-2}}{dt^{n-2}} + D(t) \frac{d^{n-1}}{dt^{n-1}}\rt) w_1^0(\tx(t),t) \nonumber \\
			= &&\frac{d^n}{dt^n}\lt(\tilde{Z}_j(t)\zeta_{1j}^0(\tx(t),t)\rt) - \tilde{Z}_{j}(t)\frac{d^n}{dt^n}\zeta_{1 j}^0(\tx(t),t) + \frac{d^n}{dt^n} \lt(\tilde{B}(t) w_1^0(\tx(t),t)\rt) \nonumber\\
			&&- \tilde{B}(t) \frac{d^n}{dt^n} w_1^0 	(\tx(t),t), \label{e35}
		\eea
		where
		\bea
			&& \tilde{Z}_j(t) := \lt(\int_0^t \dd\tau\,A_i(\tx(\tau))\mathbf{U}_{ij}(0,\tau)\rt)\\
			&& \tilde{\zeta}_{1j}^0(\tx(t),t) := \mathbf{U}_{jk}(t,0) \frac{\p}{\p y_k(t)} w_1^0 (\tx(t),t)
		\eea
		and
		\be \tilde{B}(t) := \int_0^t \dd\tau \; B(\tx(\tau)). \ee
		Combining eqs. (\ref{e33}, \ref{e34}, \ref{e35}) we obtain
		\bea
			\overline{V}_{NLL} (\tx (t),t) &=& \phi^4 \sum_{k=0}^\infty \frac{1}{k!}\lt( l\rt)^k\lt[ \lt(\frac{d}{dt}\rt)^k w_2^0(\tx(t),t) + \lt(\frac{d}{dt}\rt)^k\lt(\tilde{Z}_j(t)\zeta_{1j}^0 (\tx(t),t)\rt)\rt.\nn\\
			&& - \tilde{Z}_j(t)\lt(\frac{d}{dt}\rt)^k\zeta_{1j}^0(\tx(t),t) + \lt(\frac{d}{dt}\rt)^k\lt(\tilde{B}(t)w_{1}^0 (\tx(t),t)\rt) \nn\\
			&& \lt. - \tilde{B}(t)\lt(\frac{d}{dt}\rt)^kw_{1}^0(\tx(t),t)\rt] \nn\\
			&=& \phi^4 \lt[w_2^0 \lt(\tx \lt(t + l\rt), t + l\rt) + \lt(\tilde{Z}_j\lt(t + l\rt) - \tilde{Z}_j(t)\rt) \zeta_{1j}^0
			\lt(\tx \lt(t + l\rt), t + l\rt)\rt.\nonumber\\
			&& \lt. + \lt(\tilde{B}\lt(t + l\rt) - \tilde{B}(t)\rt) w_{1}^0
			\lt(\tx \lt(t + l\rt), t + l\rt)\rt] \nn\\
			\Rightarrow V_{NLL} = V_{NLL}(\tx(0), 0) &=& \phi^4 \lt[ w_2^0 \lt(\tx \lt( l\rt), l\rt) + \tilde{Z}_j\lt( l\rt)\zeta_{1j}^0 \lt(\tx \lt( l\rt), l\rt) + \tilde{B}\lt( l\rt) w_1^0 \lt(\tx \lt( l\rt), l\rt)\rt] \nn\\
			&=& \phi^4 K \Bigg\{ p_2^0 \lt(\tx \lt( l\rt)\rt) + \int_0^l d\tau\lt[ A_i(\tx(\tau)) \mathbf{U}_{ij}(0,\tau)\rt].\lt[ \mathbf{U}_{jk}\lt( l ,0\rt)\frac{\partial}{\partial y_k( l)} p_1^0 \lt(\tx \lt( l\rt)\rt)\rt] \nn\\
			&& +\,\,\, 4 \lt.\int_0^l d\tau \lt[\gamma_1^2 (\tx (\tau)) - \gamma_2(\tx (\tau))\rt] p_1^0\lt(\tx \lt( l\rt)\rt)\rt\rbrace
		\label{e36}\eea
		with $A_i(\tx(\tau))$'s given by (\ref{e29}). These terms can be calculated in closed form using the expression of
		\textbf{U} given by (\ref{e32}, \ref{e27}b).

		We now investigate the power series form of $V_{NLL}$, when expressed as a series in $(t-w)$, like the characteristic
		functions. We use the previously introduced notations to represent the series. We express all the coupling dependent
		terms in (\ref{e36}) as power series using the notation and collect similar terms to express $V_{NLL}$ as a
		power series. The result is:
		\be
			K^{-1} \phi^{-4} V_{NLL} = [2]_{-2}^l\, \mathrm{ln}\lt(1-lw^{-1}\rt) + [0]_{0}^0[2]_{-2}^l + [1]_{-1}^0[1]_{-1}^l \label{e53}
		\ee
		\vspace{0.2cm}
		
	  We note that, the calculations done so far are compatible with the results previously obtained for $\phi^4$ theory
		which is the limit of MSED at $\alpha \rightarrow 0$. We observe that in the limit $\alpha \to 0$ the expression
		(\ref{e36}) reduces to:
		\bea
			V_{NLL} &=& \phi^4 \lt[ -\frac{25}{12} b_2 \tg\lt(l\rt)^2 + \int_0^l \dd\tau [\beta_3(\tg(\tau))(1-b_2g\tau)^2] \lt(1-b_2gl\rt)^{-2} \frac{\p\tg}{\p\tg} - 4\int_0^l \dd\tau \gamma_2(\tg(\tau)) \tg\lt(l\rt) \rt] \nn\\
			&=& \phi^4g^2 \lt[ -\frac{25}{12} \frac{b_2}{\lt(1-b_2gl\rt)^2} - \frac{b_3}{b_2} \frac{\mathrm{ln} \lt(1-b_2gl\rt)}{ \lt(1-b_2gl\rt)^2 } - \frac{4g_2}{b_2} \frac{1}{1-b_2gl} + \frac{4g_2}{b_2} \frac{1}{\lt(1-b_2gl\rt)^2}\rt] \nn
		\eea
		which is indeed the correct single coupling $V_{NLL}$ as evaluated in ref. \cite{unique}.

\section{ A RECURSIVE RELATION FOR LEADING LOGARITHM CONTRIBUTIONS TO $V$} \label{secSING}
	\subsection{A generic expansion of the effective potential} \label{secGEP}
		From ref. \cite{unique}, we see that in $\phi^4$ theory,
		\be V_{N^{P-1}LL} = \phi^4g^P \sum_{i=1}^P \sum_{j=0}^{i-1} c_{i,j} \frac{(\ln (1-b_2gl))^j}{(1-b_2gl)^i} \label{e70}\ee
		In our model $V_{N^{P-1}LL}$ must reduce to (\ref{e70}) in the limit $\alpha \rightarrow 0$. But before proceeding 
		with our ansatz about the general form of $V_{N^{P-1}LL}$ in MSED we define,
		\be\begin{split} \chi(\alpha, g) &:= \ta(w) = \alpha\,f_1\lt(\frac{\alpha}{g}\rt) \\
			h(\alpha, g) &:= \frac{1}{b_{2,0}^g w} = \frac{16\pi^2}{5w} = g\,f_2\lt(\frac{\alpha}{g}\rt)
		\end{split}\label{e56}\ee
		where both the functions $f_1(x)$ and $f_2(x)$ are of $\mathcal{O}(x^0)$. In fact, looking at the definition of $\ta(t)$ (eq. (\ref{alpha_t})) and the limit of $w$ (eq. (\ref{limit_w})) we observe that,
		\be
			f_1(x) = 1 + \mathcal{O}(x) \qquad \mbox{and,} \qquad f_2(x) = 1 + \mathcal{O}(x) \label{f1_f2_order_1}
		\ee
		We also define,
		\be
			u := 1-lw^{-1} = 1- b_{2,0}^ghl := 1-\frac{b_{2,0}^g}{2} \xi;\quad [\xi = Lh] \label{e90}
		\ee
		For future reference, the inverse of (\ref{e56}) is of the form:
		\be\begin{split} \alpha(\chi, h) &= \chi\,\bar f_1\lt(\frac{\chi}{h}\rt) \\
			g(\chi, h) &= h\,\bar f_2\lt(\frac{\chi}{h}\rt) 
		\end{split}\label{e73}\ee
		where $\bar f_1(x)$ and $\bar f_2(x)$ both are of the form $1 + \mathcal{O}(x)$.
		
		Now, based on the limiting expression (\ref{e70}) and equations (\ref{e71}, \ref{e53}) we make the following ansatz:
		\be K^{-1} \phi^{-4} V_{N^{P-1}LL} = \sum_{n=1}^P\sum_{m=0}^{n-1} a_{n,m}[P-n]_{n-P}^0[n]_{-n}^l(\mathrm{ln}\,u)^m \nn\ee
		We simplify this expression,
		\be
			[P-n]_{n-P}^0[n]_{-n}^l = \sum_{N=0}^{\infty} \chi^N w^{N-P} \lt(\sum_{i=-n}^{N-n} c_{i,N}^P u^i\rt) \label{simplifyprod}
		\ee
		Therefore, \be
			K^{-1} \phi^{-4} V_{N^PLL} = \sum_{N=0}^{\infty} \chi^N h^{P+1-N} \lt[\sum_{i=1-N}^{P+1}\sum_{m=0}^{min(N-1+i,P)} \s_{i,m,N}^P \frac{(\mathrm{ln}\,u)^m}{u^i}\rt] \label{e63}
		\ee
		We define,
		\be S_{N,P}(\xi) = \lt[\sum_{i=1-N}^{P+1}\sum_{m=0}^{min(N-1+i,P)} \s_{i,m,N}^P \frac{(\mathrm{ln}\,u)^m}{u^i}\rt] \label{e72}\ee
		Now the effective potential $V$ becomes,
		\be
			V = K \phi^4 \sum_{P=0}^{\infty} \sum_{N=0}^{\infty} \chi^N h^{P+1-N} S_{N,P}(\xi)
		\label{e74}\ee

	\subsection{Determination of the recurrence relation for finding $V$} \label{secREC}
		Now we use the RG equation to determine the coefficients $\s_{i,m,N}^P$ in (\ref{e72}) which will fix the effective
		potential. In the following calculation we consider $\chi$ and $h$ as independent variables in stead of $\alpha$ and $g$.
		
		From equation (\ref{e6}) we have:
		\bea
			%\lt(\mu\frac{\p}{\p\mu} + \beta^g\frac{\p}{\p g} + \beta^{\alpha}\frac{\p}{\p\alpha} + \gamma\phi\frac{\p}{\p\phi}\rt) V &=& 0 \nn\\\Rightarrow 
			\lt(\mu\frac{\p}{\p\mu} + \lt(\beta^g \frac{\p \chi}{\p g} + \beta^{\alpha} \frac{\p \chi}{\p \alpha}\rt) \frac{\p}{\p \chi} + \lt(\beta^g \frac{\p h}{\p g} + \beta^{\alpha} \frac{\p h}{\p \alpha}\rt) \frac{\p}{\p h} + \gamma\phi\frac{\p}{\p\phi}\rt) V &=& 0 \label{e75}
		\eea
		We now want to find a recurrence relation for the $\s$'s that will be equivalent to the above PDE. To that end we substitute $V$ from (\ref{e74}) in (\ref{e75}), expand everything as a power series in $\chi$, $h$, $\la$ and $u$ and equate the $\mathcal{O}(\chi^mh^{q+2-m}u^{-j}\la^n)$ term to zero. This is a fairly straightforward albeit tedious process. We shall only write the final result. In the process we write the beta functions as functions of $\chi$ and $h$.
		\begin{subequations}\label{e85}\bea
			\beta^g &=& \sum_{n=2}^{\infty} \sum_{i=0}^{\infty} \bar b_{n,i}^g h^{n-i}\chi^i \\
			\beta^{\alpha} &=& \sum_{n=2}^{\infty} \sum_{i=0}^{\infty} \bar b_{n,i+1}^{\alpha} h^{n-i-1}\chi^{i+1}
		\eea\end{subequations}
		(\ref{e85}) defines $\bar b_{n,i}^g$ and $\bar b_{n,i}^\alpha$. Referring to (\ref{e73}) we note,
		\be \frac{\p \chi}{\p g} = \mathcal{O}\lt(\frac{\chi}{h}\rt);\quad \frac{\p \chi}{\p \alpha} = \mathcal{O}\lt(\lt(\frac{\chi}{h}\rt)^0\rt);\quad \frac{\p h}{\p g} = \mathcal{O}\lt(\lt(\frac{\chi}{h}\rt)^0\rt);\quad \frac{\p h}{\p \alpha} = \mathcal{O}\lt(\lt(\frac{\chi}{h}\rt)^{-1}\rt) \nn\ee
		This implies,
		\begin{subequations}\bea \beta^g \frac{\p \chi}{\p g} + \beta^{\alpha} \frac{\p \chi}{\p \alpha} &=& \sum_{n=2}^{\infty} \sum_{i=0}^{\infty} k_{n-i-1,i+1}^1 h^{n-i-1}\chi^{i+1} \\
		\quad \beta^g \frac{\p h}{\p g} + \beta^{\alpha} \frac{\p h}{\p \alpha} &=& \sum_{n=2}^{\infty} \sum_{i=0}^{\infty} k_{n-i,i}^2 h^{n-i}\chi^i \\
		\gamma &=& \sum_{n=1}^{\infty} \sum_{i=0}^{\infty} k_{n-i,i}^3 h^{n-i}\chi^i
		\eea\label{e62}\end{subequations}
		(\ref{e62}) defines $k_{i,j}^1$, $k_{i,j}^2$ and $k_{i,j}^3$. Furthermore, we transform the partial derivatives appearing in (\ref{e75}) as follows:
		\be\begin{split}
			\frac{\p}{\p \phi} &\rightarrow \frac{\p}{\p \phi} + \frac{\p \xi}{\p \phi} \frac{\p}{\p \xi} = \frac{\p}{\p \phi} + 2h\phi^{-1}\frac{\p}{\p \xi} \\
			\frac{\p}{\p h} &\rightarrow \frac{\p}{\p h} + \frac{\p \xi}{\p h} \frac{\p}{\p \xi} = \frac{\p}{\p h} + h^{-1}\xi \frac{\p}{\p \xi} \\\
		\frac{\p}{\p \mu} &\rightarrow \frac{\p \xi}{\p \mu} \frac{\p}{\p \xi} = -2h\mu^{-1}\frac{\p}{\p \xi}
		\end{split}\label{chain}\ee
		We also need to use the inverse of $\alpha$ from (\ref{aginch}),
		\be
			\alpha^{-1} = \chi^{-1} \sum_{i=0}^\infty A'_i \chi^i h^{-i} \label{a_inverse}
		\ee
		which defines $A'_i$. In terms of these newly defined constants, the $\mathcal{O}(\chi^mh^{q+2-m}u^{-j}\la^n)$ terms of (\ref{e75}) becomes:
		\bea
			&& -(j-1)\kk \s_{j-1,n,m}^q + (n+1) \kk \s_{j-1,n+1,m}^q + \sum_{P=0}^q \sum_{i=0}^m \Big[k_{q+1-P-i,i+1}^1 (m-i) \s_{j,n,m-i}^P \nn\\
			&& + k_{q+2-P-i,i}^2 \{(P+1-m+i-j) \s_{j,n,m-i}^P +(j-1)\s_{j-1,n,m-i}^P \nn\\ && - (n+1) \s_{j-1,n+1,m-i}^P + (n+1) \s_{j,n+1,m-i}^P\} + 4k_{q+1-P-i,i}^3 \s_{j,n,m-i}^P\Big] \nn\\&& + \sum_{p=0}^{q-1} \sum_{i=0}^m k_{q-P-i,i}^3 \lt[(j-1)\kk \s_{j-1,n,m-i}^P - (n+1) \kk \s_{j-1,n+1,m-i}^P\rt] \nn\\
			&& + 4\tau_{0,1} \sum_{i=0}^{m-1} \lt(\frac{-b_{0,2}^\alpha}{b_{2,0}^g}\rt)^a \lt[ \sum_{c=i}^{m-1} \lt( \sum_{t=2}^{q+2} \sum_{a=0}^{c-i} \frac{A'_{c-i-a}}{b_{2,0}^g} \bar b_{t,a+1}^\alpha \lt(\s_{j+i,n,m-1-c}^{q+2-t} - \s_{j+i+1,n,m-1-c}^{q+2-t}\rt) \rt.\rt. \nn\\
			&& \lt.\lt. + \sum_{t=1}^q k_{t-c+i,c-i}^3 \s_{j+i,n,m-1-c}^{q-t} \rt) - \s_{j+i,n,m-1-i}^q \rt] = 0 \nn\\
			&\mathrm{or,}& R(j,n,m,q) = 0 \label{master_rec}
		\eea
		We shall use this as a recursive definition of $\s_{j,n+1,m}^q$. $\s_{i,j,k}^l$ can be considered as a function $\s(i,j,k,l)$ defined on a 4D integer lattice $L \subset \mathbb{Z}^4$. We recall from the definition of $\s$ that $\s_{i,j,k}^l = 0$ when one or more of the following seven conditions are satisfied: $i < 1-k$, $i > l+1$, $j < 0$, $j > k-1+i$, $j > l$, $k < 0$, and $l < 0$. So we can take our lattice to be defined as:
		\be
			L = \{ (i,j,k,l) \in \mathbb{Z}^4\; |\; k \ge 0,\, l \ge j \ge 0,\, l \ge i-1 \ge j-k\} \label{lattice}
		\ee
		$\s(\mb{n}) = 0$ for any $\mb{n} \notin L$. Now, the recurrence (\ref{master_rec}) allows us to write $\s(\mb{n})$, for any $\mb{n} \in L$, in terms of $\s(\mb{m})$ where $\mb{m}$ belongs to the following set:
		\be\begin{split}
			B_0 &= \{(i,j,k,l) \in L\; |\; j = 0\} %\\
			%B_1 &= \{(i,j,k,l) \in L\; |\; i-j+k-1 = 0\}, \\
			%B_2 &= \{(i,j,k,l) \in L\; |\; i-l-1=0\}. \nn
		\end{split}\label{master_bound_4}\ee
		%We define the boundary $\mathcal{B}$ to be the union of these sets, i.e. $\mathcal{B} := \bigcup_{i=0}^2 B_i$.

\section{SINGULARITY ANALYSIS} \label{sec_singularity_analysis}
\subsection{Some general constructions for recurrence relations}
	Before proceeding with the investigation of the singularity structure of the effective potential we find some general results 
	about recurrence relations that will be useful later. We start with a generic recurrence relation in one variable. We consider 
	the following relation which defines $X(n)$ recursively:
	\be
		X(n) = \sum_{i=0}^{n-1} f(n,i)\, X(i) \label{gen_1d_rec}
	\ee
	with $f$ not identically zero but $f(n, m) = 0$ whenever $n \le m$. We shall say that the point $n$ depends recursively on the point $i$ if $f(n, i) \ne 0$. Now, we want to express $X(n)$ in terms of $X(0)$. We can write down the first few cases easily:
	\bea
		X(1) &=& f(1,0) X(0) \label{X(1)}\\
		X(2) &=& f(2,0) X(0) + f(2,1) X(1) \nn\\
		&=& [f(2,0) + f(2,1) f(1,0)]\, X(0)\quad [\mbox{Using } (\ref{X(1)})] \label{X(2)}\\
		X(3) &=& f(3,0) X(0) + f(3,1) X(1) + f(3,2) X(2) \nn\\
		&=& [f(3,0) + f(3,1) f(1,0) + f(3,2) f(2,0) + f(3,2) f(2,1) f(1,0)]\, X(0) \label{X(3)}\quad [\mbox{Using } (\ref{X(1)}) \mbox{ and } (\ref{X(2)})] \nn\\
		&& \mbox{and so on...} \nn
	\eea
	In fact, induction gives the following expression for $X(n)$:
	\be
		X(n) = \sum_{s=1}^n\;\; \sum_{0 < p_1 < \cdots < p_{s-1} < n}\;\; \prod_{i=1}^s f(p_i, p_{i-1})\, X(0) \label{X(n)}
	\ee
	with $p_0 = 0$ and $p_s = n$. Now we define a path $\mathcal{P}(n,m)$ connecting two nonnegative integers $n$ and $m$ (we assume, $m < n$) to be an ordered collection of $k+1$ points of $\mathbb{Z}_{\ge 0}$ denoted $\mathcal{P}(n,m) = (\mathcal{P}^0, \mathcal{P}^1, \cdots, \mathcal{P}^k)$ where $f(\mathcal{P}^{i+1}, \mathcal{P}^i) \ne 0$ with $\mathcal{P}^0 = m$ and $\mathcal{P}^k = n$. The length of the path, denoted $|\mathcal{P}(n,m)|$, is $k$. For any two non negative integers $m$ and $n$ with $m < n$ if we write $S(n,m)$ for the set of all paths connecting $n$ and $m$ then we define the functions $\mathcal{F}$ and $\mathcal{T}$ as:
	\be
		\mbox{For any } \mathcal{P} \in S(n, m),\quad \mathcal{F}(\mathcal{P}) := \prod_{i=1}^{|\mathcal{P}|} f(\mathcal{P}^i, \mathcal{P}^{i-1}) \qquad \mbox{and,}\qquad \mathcal{T}(n,m) := \sum_{\mathcal{P} \in S(n,m)} \mathcal{F}(\mathcal{P}) \label{transition_1d}
	\ee
	In our present case we notice that:
	\be
		\sum_{\mathcal{P} \in S(n,m)} := \sum_{s=1}^{n-m}\;\; \sum_{m < \mathcal{P}^1 < \cdots < \mathcal{P}^{s-1} < n}
	\ee
	with $\mathcal{P}^0 = m$ and $\mathcal{P}^s = n$. Therefore using the function $\mathcal{T}$ we can rewrite (\ref{X(n)}) as
	\be
		X(n) = \mathcal{T}(n,0) X(0) \label{XT_1d}
	\ee
	%\vspace{0.5cm}
	
	Now we move on to recurrence relations with more that one variable. Suppose $L \subset \mathbb{Z}^N$ is the lattice on which the function $X$ is defined and $X(\mb{n})$ is defined by the following recurrence relation:
	\be
		X(\mb{n}) = \sum_{\mb{i} \in M(\mb{n})} f(\mb{n}, \mb{i})\, X(\mb{i}) \label{gen_Nd_rec}
	\ee
	where $M(\mb{n})$ is a finite subset of $L$ that depends on $\mb{n}$, a boldface letter represents an element of $L$ and $f$ is not identically zero but $f(\mb{n}, \mb{m}) = 0$ if $X(\mb{m})$ does not appear in the recursive definition of $X(\mb{n})$. As before, an ordered collection of points of $L$ denoted by $\mathcal{P}(\mb{n}, \mb{m}) = (\mathcal{P}^0, \mathcal{P}^1, \cdots, \mathcal{P}^k)$ where $f(\mathcal{P}^{i+1}, \mathcal{P}^i) \ne 0$ with $\mathcal{P}^0 = \mb{m}$ and $\mathcal{P}^k = \mb{n}$, will be called a path of length $k$ (denoted by $|\mathcal{P}(\mb{n}, \mb{m})|$) connecting $\mb{n}$ and $\mb{m}$. And if $S(\mb{n}, \mb{m})$ denotes the set of all paths connecting $\mb{n}$ and $\mb{m}$ then we define 
	\be
		\mbox{For any } \mathcal{P} \in S(\mb{m}, \mb{n}),\quad \mathcal{F}(\mathcal{P}) := \prod_{i=1}^{|\mathcal{P}|} f(\mathcal{P}^i, \mathcal{P}^{i-1}) \quad \mbox{and,} \quad \mathcal{T}(\mb{n},\mb{m}) := \sum_{\mathcal{P} \in S(\mb{n},\mb{m})} \mathcal{F}(\mathcal{P}) \label{transition_Nd}
	\ee
	
	For a recurrence relation with $N$ variables the boundary $\mathcal{B} \subset L$ will in general consist of $N-1$ dimensional sub lattices of $L$ i.e. there will exist a decomposition of $\mathcal{B}$, $\mathcal{B} = \bigcup_i B_i$, such that $B_i \subset \mathbb{Z}^{N-1}$ and the generalized form of (\ref{XT_1d}) is:
	\be
		X(\mb{n}) = \sum_{\mb{m} \in \mathcal{B}} \mathcal{T}(\mb{n}, \mb{m}) X(\mb{m}) \label{XT}
	\ee

\subsection{More constructions}
	We recall that in our case we have a four dimensional integer lattice $L$ (defined in (\ref{lattice})) and we write the elements of $L$ as $\mb{m} = (m_1, m_2, m_3, m_4)$. We now define an equivalence relation for the elements of $L$. We shall say that two elements $\mb{m}, \mb{n} \in L$ are equivalent and write $\mb{m} \sim \mb{n}$ if $m_1 - n_1 = m_2 - n_2 = m_4 - n_4$ and $m_3 = n_3$. Thus, for any $(i, j, m, n) \in L$ we have the following chain of equivalencies:
	\be
		B_0 \ni (i-j, 0, m, n-j) \sim (i-j+1, 1, m, n-j+1) \sim \cdots \sim (i, j, m, n) \sim (i+1, j+1, m, n+1) \sim \cdots \nn
	\ee
	The fact that such a chain of equivalencies will always be bounded from below by an element of $B_0$ can easily be seen from the constraints that define $L$ in (\ref{lattice}). We note that an element of $L$ is equivalent to exactly one element of $B_0$. Two distinct elements of $B_0$ are obviously inequivalent. Also, for each element of $B_0$ we can construct a chain of equivalencies starting from that element. Hence we see that,
	\be
		\widetilde L := L/\sim \;\, \cong B_0
	\ee
	This allows us to construct the following bijection:
	\be\begin{split}
		\vartheta &: B_0 \times \mathbb{Z}_{\ge 0} \rightarrow L \\
		\vartheta &: ((a, 0, m, b), n) \mapsto (a+n, n, m, b+n)
	\end{split}\label{vartheta}\ee
	We shall sometimes write $\vartheta((a, 0, m, b), n)$ as $\vartheta_{a,m}^b(n)$ for short. Now, using this function we partition the effective potential $V$ as follows:
	\be\begin{split}
		\mbox{First we define, for any } \mb{c} \in B_0, \quad V_{\mb{c}} &:= \sum_{n=0}^\infty \s(\vartheta(\mb{c}, n)) \eta^n; \quad \lt[\eta := h \frac{\ln u}{u} \rt] \\
		\mbox{Then } V \mbox{ becomes, } \quad V &= K h \phi^4 \sum_{\mb{c} \in B_0} \chi^{c_3} V_{\mb{c}} \frac{(\ln u)^{-c_4}}{u^{c_1-c_4}}
	\end{split}\label{V_part}\ee
	We shall look for the singularities of $V_{\mb{c}}$.
	
	We need to write the recurrence relation (\ref{master_rec}) in a more managable form. To that end, we define the functions $T_1$, $T_2$, $T_3$ and $T_4$ as follows:
	\be\begin{split}
		T_i &: L \rightarrow L \\
		T_i &: (n_1, n_2, n_3, n_4) \mapsto (n_1 - \delta_{i,1}, n_2 - \delta_{i,2}, n_3 - \delta_{i,3}, n_4 - \delta_{i,4})
	\end{split}\label{T}\ee
	These functions have well defined inverses except at some boundaries of $L$ and they obviously commute, i.e., $T_i \circ T_j = T_j \circ T_i$ for any $1 \ge i, j \ge 4$. Now, if we take (\ref{master_rec}) to be the recursive definition of $\s(\vartheta(\mb{c}, k))$ for some $\mb{c} \in B_0$ and $k \ge 0$ then we can write it in the generic form:
	\be
		\s(\vartheta(\mb{c}, k)) = \sum_{(s_1, s_2, s_3, s_4) \in \mathcal{I}(\mb{c}, k)} \psi\lt(\mb{c}, k, \{s_i\}_{i=1}^4\rt) \prod_{i=1}^4 T_i^{s_i} \big(\vartheta(\mb{c}, k)\big) \label{master_rec_L}
	\ee
	where in the terminology of (\ref{gen_Nd_rec}):
	\be
		f\lt(\vartheta(\mb{c}, k), \prod_{i=1}^4 T_i^{s_i} \big(\vartheta(\mb{c}, k)\big)\rt) = \psi\lt(\mb{c}, k, \{s_i\}_{i=1}^4\rt) \label{f_psi}
	\ee
	and $\mathcal{I}(\mb{c}, k)$ is a set of 4-tuples such that:
	\be
		M(\vartheta(\mb{c}, k)) = \lt\{ \prod_{i=1}^4 T_i^{s_i} \big(\vartheta(\mb{c}, k)\big)\; |\; (s_1, s_2, s_3, s_4) \in \mathcal{I}(\mb{c}, k) \rt\}
	\ee
	In particular $s_3$ takes value from the set $\{0, 1, \cdots, c_3\}$. Now let, $\mathcal{P} = (\mathcal{P}^0, \mathcal{P}^1, \cdots, \mathcal{P}^N = \vartheta(\mb{c}, k))$ be a path connecting $\vartheta(\mb{c}, k)$ and $\mathcal{P}^0$, then:
	\bea
		\mathcal{F}(\mathcal{P}) &=& \prod_{i=1}^N f(\mathcal{P}^i, \mathcal{P}^{i-1}) \nn\\
		&=& \prod_{i=1}^N \psi\lt(\mb{c}_i, k_i, \{s_{j,i}\}_{j=1}^4\rt) \label{F(P)}
	\eea
	where $(\mb{c}_N, k_N) = (\mb{c}, k)$ and,
	\be
		(\mb{c}_{i-1}, k_{i-1}) = \vartheta^{-1}\lt( \prod_{j=1}^4 T_j^{s_{j,i}} (\vartheta(\mb{c}_i, k_i)) \rt) \label{temp_0}
	\ee
	Iteration of (\ref{temp_0}) gives,
	\be
		\vartheta(\mb{c}_0, k_0) = \prod_{j=1}^4 T_j^{\sum_{i=1}^N s_{j,i}} (\vartheta(\mb{c}, k))
	\ee
	This shows in particular that if we write $\vartheta(\mb{c}, k) = (a_1, a_2, a_3, a_4)$ and $\vartheta(\mb{c}_0, k_0) = (b_1, b_2, b_3, b_4)$ then,
	\be
		b_3 = a_3 - \sum_{i=1}^N s_{3,i}
	\ee
	However, $(b_1, b_2, b_3, b_4) \in L$ only if $b_3 \ge 0$. Therefore if we define the set $I := \{i \in \{1, 2, \cdots, N\}\; |\; s_{3,i} \ge 1\}$ then we get,
	\begin{align}
		b_3 &= a_3 - \sum_{i \in I} s_{3,i} \ge 0 \nn\\
		\Rightarrow |I| &= \sum_{i \in I} 1 \le \sum_{i \in I} s_{3,i} \le a_3 \nn
	\end{align}
	Now, if for all $i \in I$, $\psi\lt(\mb{c}, k, \{s_{j,i}\}_{j=1}^4\rt)$ is at most linear in the elements of a set of symbols $\Omega$ then from (\ref{F(P)}) we see that $\mathcal{F}(\mathcal{P})$ is a polynomial in the elements of $\Omega$ of degree at most $|I| \le a_3$. From (\ref{master_rec}) we see that if we consider it as a recursive definition of $\s(j, n+1, m, q)$ then the coefficients of terms of the form $\s(\_,\_,m-i,\_)$ with $i \ge 1$ are homogenously linear in $k_{\_,j+1}^1$, $k_{\_,j}^2$, $k_{\_,j}^3$ with $j \ge 1$ and $\tau_{0,1}$. Thus, for this case we can take $\Omega$ to be defined as
	\be
		\Omega = \{k_{\_,j+1}^1, k_{\_,j}^2, k_{\_,j}^3, \tau_{0,1}\; |\; j \ge 1\}. \label{Omega}
	\ee
	Then the above discussion shows that for any path $\mathcal{P}$ connecting $\s(\vartheta(\mb{c}, n))$ and $\s(\mb{m})$ by (\ref{master_rec}), $\mathcal{F}(\mathcal{P})$ and by extension $\mathcal{T}(\vartheta(\mb{c}, n), \mb{m})$ is a polynomial in the elements of $\Omega$ of degree at most $\vartheta(\mb{c}, n)_3 = c_3$, for any $n \ge 0$. This allows for a great simplification. Using (\ref{V_part}) we can write $V_{\mb{c}}$ as a finite degree polynomial in the elements of $\Omega$, the degree being bounded by $c_3$. Thus if $V_{\mb{c}}$ is finite for some particular value of the elements of $\Omega$ then $V_{\mb{c}}$ is finite for any value of the elements of $\Omega$. This implies, if $\eta = \omega$ is a singularity of $V_{\mb{c}}$ then $\omega$ is independent of the elements of $\Omega$. Therefore, without any loss of generality, we shall take the elements of $\Omega$ to be zero and calculate the singularities of $V_{\mb{c}}$.
	
	Now, if we equate the elements of $\Omega$ (defined in (\ref{Omega})) to zero, then (\ref{master_rec}) reduces to:
	\bea
		&& -(j-1)\kk \s_{j-1,n,m}^q + (n+1) \kk \s_{j-1,n+1,m}^q + \sum_{P=0}^q \Big[k_{q+1-P,1}^1 m \s_{j,n,m}^P \nn\\
		&& + k_{q+2-P,0}^2 \{(P+1-m-j) \s_{j,n,m}^P +(j-1)\s_{j-1,n,m}^P \nn\\
		&& - (n+1) \s_{j-1,n+1,m}^P + (n+1) \s_{j,n+1,m}^P\} + 4k_{q+1-P,0}^3 \s_{j,n,m}^P\Big] \nn\\
		&& + \sum_{p=0}^{q-1} k_{q-P,0}^3 \lt[(j-1)\kk \s_{j-1,n,m}^P - (n+1) \kk \s_{j-1,n+1,m}^P\rt] = 0 \nn\\
		&\mathrm{or,}& R'(j,n,m,q) = 0 \label{master_rec_1}
	\eea
	Instead of directly trying to solve it for $\s_{j, n+1, m}^q$ we shall consider it as a recursive definition of $\s_{j, n, m}^q$ and write $\s_{j, n, m}^q$ in terms of elements of $\s(B_1)$ and $\s(B_2)$ where $B_1$ and $B_2$ are defined as follows:
	\be\begin{split}
		B_1 &:= \{(i,j,k,l) \in L\; |\; i-j+k-1 = 0\}, \\
		B_2 &:= \{(i,j,k,l) \in L\; |\; i-l-1=0\}% \\
		%\mbox{We also define,}\quad B_{12} &:= B_1 \cup B_2
	\end{split}\label{pseudo_boundary}\ee
	Then we shall write the elements of $\s(B_1 \cup B_2)$ in terms of elements of $\s(B_0)$ which will give us $\s(j, n, m, q)$ in terms of the elements of $\s(B_0)$. The reason for doing this is that it will turn out to be much easier to relate the elements of $\s(B_1 \cup B_2)$ to the elements of $\s(B_0)$ than it is with an arbitrary element of $\s(L)$. Now, using (\ref{master_rec_1}), in accordance with (\ref{XT}), for any $\mb{c} \in B_0$ and $n \ge 0$ we can write:
	\bea
		\s(\vartheta(\mb{c}, n)) &=& \sum_{\mb{m} \in B_1 \cup B_2} \mathcal{T}_{(\ref{master_rec_1})}(\vartheta(\mb{c}, n), \mb{m}) \s(\mb{m}) \nn\\
		&=& \lt( \sum_{\mb{m} \in B_1} + \sum_{\mb{m} \in B_2 \backslash (B_1 \cap B_2)} \rt) \mathcal{T}_{(\ref{master_rec_1})}(\vartheta(\mb{c}, n), \mb{m}) \s(\mb{m}) \label{L_to_B1_B2}
	\eea
	The subscript (\ref{master_rec_1}) of $\mathcal{T}$ means that the function is to be evaluated from equation (\ref{master_rec_1}). Now, any element of $B_1$ can be written as $(n+1-m, n, m, n+i)$. From (\ref{master_rec_1}) we see,
	\bea
		&& R'(n+1-m, n, m, n+i) = 0 \nn\\
		&\Rightarrow& \sum_{p = n}^{n+i} \Big[k_{n+i+1-p,1}^1 m + k_{n+i+2-p,0}^2 (p-n) + 4k_{n+i+1-p,0}^3\Big] \s_{n+1-m,n,m}^p = 0 \label{rec_B1}
	\eea
	We note that $(n+1-m, n, m, P) \in B_1$ and hence (\ref{rec_B1}) can be considered as a standalone recursive definition of the elements of $B_1$. This is a simple 1D recurrence of the form (\ref{gen_1d_rec}). Therefore in accordance with (\ref{XT_1d}) we can write:
	\be
		\s_{n+1-m, n, m}^{n+i} = \mathcal{T}_{(\ref{rec_B1})}((n+1-m, n, m, n+i), (n+1-m, n, m, n)) \s_{n+1-m, n, m}^n \label{B1_to_B3}
	\ee
	We note that $\mathcal{T}_{(\ref{rec_B1})}((n+1-m, n, m, n+i), (n+1-m, n', m, n')) = 0$ for any $n' \ne n$. We define the set $B_3$ for elements like $(n+1-m, n, m, n)$:
	\be\begin{split}
		B_3 &:= \{\mb{m} \in B_1\; |\; m_2 = m_4\} \\
		\mbox{We also define, } \quad B_4 &:= B_2 \backslash (B_1 \cap B_2)
	\end{split}\label{B34}\ee
	Now, (\ref{L_to_B1_B2}) becomes:
	\be
		\s(\vartheta(\mb{c}, n)) = \sum_{\mb{k} \in B_3} \sum_{\mb{m} \in B_1} \mathcal{T}_{(\ref{master_rec_1})}(\vartheta(\mb{c}, n), \mb{m}) \mathcal{T}_{(\ref{rec_B1})}(\mb{m}, \mb{k}) \s(\mb{k}) + \sum_{\mb{m} \in B_4} \mathcal{T}_{(\ref{master_rec_1})}(\vartheta(\mb{c}, n), \mb{m}) \s(\mb{m}) \label{L_to_B1_B2-1}
	\ee
	We now define $B_{34} := B_3 \cup B_4$ and for any $\mb{c} \in B_0$, $n \ge 0$, $\mb{m} \in B_{34}$ we define the following function:
	\be
		\mathcal{T}'(\vartheta(\mb{c}, n), \mb{m}) := \lt\{
		\begin{array}{ll}
			\sum_{\mb{k} \in B_1} \mathcal{T}_{(\ref{master_rec_1})}(\vartheta(\mb{c}, n), \mb{k}) \mathcal{T}_{(\ref{rec_B1})} (\mb{k}, \mb{m}) & \mbox{if } \mb{m} \in B_3 \\
			%& \\
			\mathcal{T}_{(\ref{master_rec_1})}(\vartheta(\mb{c}, n), \mb{m}) & \mbox{if } \mb{m} \in B_4
		\end{array} \rt.
	\ee
	We note that since $B_3 \cap B_4 = \emptyset$ by construction, the function is well defined. Now using this function we can rewrite (\ref{L_to_B1_B2-1}) as:
	\be
		\s(\vartheta(\mb{c}, n)) = \sum_{\mb{m} \in B_{34}} \mathcal{T}'(\vartheta(\mb{c}, n), \mb{m}) \s(\mb{m}) \label{L_to_B1_B2_2}
	\ee
	We note that any element of $B_4$ can be written as $(n+1, n-i, m, n)$ with $\{i, m\} \ne \{0\}$ and for any $j \in \mathbb{Z}$ such that $n-i+j \ge 0$ we have, $(n+1, n-i, m, n) \sim (n+1+j, n-i+j, m, n+j) \in B_4$. Thus for any element of $B_4$ the equivalence chain containing the element is completely contained in $B_4$. Therefore we see,
	\be
		\widetilde B_4 := B_4/\sim \; = B_4 \cap B_0
	\ee
	and $\vartheta : \widetilde B_4 \times \mathbb{Z}_{\ge 0} \rightarrow B_4$ is a bijection. Similar argument applies for $B_1$ and $B_3$ as well and we find that for $i \in \{1, 3\}$, $\widetilde B_i := B_i/\sim\; = B_i \cap B_0$ and $\vartheta : \widetilde B_i \times \mathbb{Z}_{\ge 0} \rightarrow B_i$ is bijective. Thus defining $\widetilde B_{34} := B_{34}/\sim$ we can rewrite (\ref{L_to_B1_B2_2}) as:
	\be
		\s(\vartheta(\mb{c}, n)) = \sum_{\mb{p} \in \widetilde B_{34}} \sum_{m=0}^\infty \mathcal{T}'(\vartheta(\mb{c}, n), \vartheta(\mb{p}, m)) \s(\vartheta(\mb{p}, m)) \label{L_to_B1_B2_3}
	\ee
	%\vspace{0.7cm}
	
\subsection{Asymptotic analysis}
	%\noindent\textbf{Asymptotic analysis}
	%\vspace{0.3cm}
	
	\noindent So far the relations we have derived are all exact. But from (\ref{V_part}) we see that the condition for convergence of $V_{\mb{c}}$ is:
	\be
		\lim_{n \to \infty} \frac{\s(\vartheta(\mb{c},n+1))}{\s(\vartheta(\mb{c},n))} \eta < 1 \label{conv_cond}
	\ee
	Thus, to determine the singularity of $V_{\mb{c}}$ we need only the asymptotic characteristics of $\s(\vartheta(\mb{c},n))$ at large $n$. We first look at the asymptotic behavior of $\s(B_4)$. From (\ref{master_rec_1}) we find (for $\{i, m\} \ne \{0\}$),
	\bea
		R'(n+1, n-i, m, n) &=& 0 \nn\\
		\Rightarrow \s_{n+1, n-i+1, m}^n &=& \frac{-1}{k_{2,0}^2 (n-i+1)} \Bigg[ \lt\{ \lt(k_{1,1}^1 - k_{2,0}^2\rt)m + 4k_{1,0}^3 \rt\} \s_{n+1,n-i,m}^n \nn\\
		&& - \lt( k_{3, 0}^2 + k_{2,0}^2 k_{1, 0}^3 \rt) \lt\{ (n-i+1) \s_{n, n-i+1, m}^{n-1} + n \s_{n, n-i, m}^{n-1} \rt\} \Bigg] \label{rec_B2}
	\eea
	If we use a bar over $\s$ to denote the asymptotic behavior, e.g., if we write $\bar\s(\vartheta(\mb{c}, n))$ for the asymptotic form of $\s(\vartheta(\mb{c}, n))$ at large $n$ then taking the large $n$ limit in (\ref{rec_B2}) we get:
	\be
		\bar\s_{n+1, n-i+1, m}^n = \frac{k_{3, 0}^2 + k_{2,0}^2 k_{1, 0}^3}{k_{2,0}^2} \lt[ \bar\s_{n, n-i+1, m}^{n-1} - \bar\s_{n, n-i, m}^{n-1} \rt] \label{rec_B2_asym}
	\ee
	The results of Appendix \ref{k_to_beta} show that $k_{1,0}^3$ is proportional to $\tau_{1,0}$ which is 0. Now, if we define $\rho := -k_{3,0}^2/k_{2,0}^2$ then, the solution to (\ref{rec_B2_asym}) is:
	\be
		\bar\s_{n+1, n-i+1, m}^n = \sum_{k=1}^i \frac{(n-k)!}{(n-i)!(i-k)!} \rho^{n+1-k} (-1)^{i-k}\, \bar\s_{k, 0, m}^{k-1}
	\ee
	Or we can write it like,
	\be
		\bar\s\lt( \vartheta_{i,m}^{i-1}(n-i+1) \rt) = \sum_{k=1}^i \mathcal{T}_{(\ref{rec_B2_asym})} \lt( \vartheta_{i,m}^{i-1}(n-i+1), \vartheta_{k,m}^{k-1}(0) \rt) \bar\s \lt( \vartheta_{k,m}^{k-1}(0) \rt) \label{sol_B2}
	\ee
	which also defines the function $\mathcal{T}_{(\ref{rec_B2_asym})} (\_,\_)$. Now,
	\be
		\lim_{n \to \infty} \frac{\mathcal{T}_{(\ref{rec_B2_asym})} \lt( \vartheta_{i,m}^{i-1}(n-i+2), \vartheta_{k,m}^{k-1}(0) \rt)}{\mathcal{T}_{(\ref{rec_B2_asym})} \lt( \vartheta_{i,m}^{i-1}(n-i+1), \vartheta_{k,m}^{k-1}(0) \rt)} = \lim_{n \to \infty} \frac{n+1-k}{n+1-i} \rho = \rho
	\ee
	Noting that any $\mb{p} \in B_0$ can be written as $\vartheta_{i,m}^{i-1}(0)$ for suitable $i$ and $m$, we find for any $\mb{p} \in \widetilde B_4 \subset B_0$, $n \ge 0$ and such $\mb{q} \in B_0$ that $\mathcal{T}_{(\ref{rec_B2_asym})} (\vartheta(\mb{p}, n), \mb{q}) \ne 0$:
	\be
		\mathcal{T}_{(\ref{rec_B2_asym})} (\vartheta(\mb{p}, n+1), \mb{q}) \sim \rho\, \mathcal{T}_{(\ref{rec_B2_asym})} (\vartheta(\mb{p}, n), \mb{q}) \quad [\mbox{as } n \to \infty] \label{sim_1}
	\ee
	
	Similarly, for the elements of $B_3$, starting from $R'(n+1-m, n-1, m, n) = 0$ and taking the large $n$ limit we get:
	\be
		\bar\s_{n+1-m, n, m}^n = \rho \bar\s_{n-m, n-1, m}^{n-1} \label{rec_B3_asym}
	\ee
	which has the solution:
	\be
		\bar\s_{n+1-m, n, m}^n = \rho^n \bar\s_{1-m, 0, m}^0 \label{sol_B3}
	\ee
	and following a similar argument as for the elements of $B_4$ we get, for any $\mb{p} \in \widetilde B_3$, $n \ge 0$ and such $\mb{q} \in B_0$ that $\mathcal{T}_{(\ref{rec_B2_asym})} (\vartheta(\mb{p}, n), \mb{q}) \ne 0$:
	\be
		\mathcal{T}_{(\ref{rec_B3_asym})} (\vartheta(\mb{p}, n+1), \mb{q}) \sim \rho\, \mathcal{T}_{(\ref{rec_B3_asym})} (\vartheta(\mb{p}, n), \mb{q}) \quad [\mbox{as } n \to \infty] \label{sim_2}
	\ee
	
	We now define the following function, for any $\mb{p} \in \widetilde B_{34}$, $n \ge 0$, and $\mb{q} \in B_0$:
	\be
		\mathcal{T}''(\vartheta(\mb{p}, n), \mb{q}) := \lt\{
		\begin{array}{ll}
			\mathcal{T}_{(\ref{rec_B3_asym})} (\vartheta(\mb{p}, n), \mb{q}) & \mbox{if } \mb{p} \in B_3 \\
			%& \\
			\mathcal{T}_{(\ref{rec_B2_asym})} (\vartheta(\mb{p}, n), \mb{q}) & \mbox{if } \mb{p} \in B_4
		\end{array} \rt. \label{T''}
	\ee
	and then we can rewrite (\ref{L_to_B1_B2_3}) as:
	\be
		\bar\s(\vartheta(\mb{c}, n)) = \sum_{\substack{ \mb{q} \in B_0 \\ \mb{p} \in \widetilde B_{34} }} \sum_{m=0}^\infty \mathcal{T}'(\vartheta(\mb{c}, n), \vartheta(\mb{p}, m)) \mathcal{T}''(\vartheta(\mb{p}, m), \mb{q}) \bar\s(\mb{q}) \label{L_to_B1_B2_4}
	\ee
	Now, for any $\mb{c}, \mb{p} \in B_0$ and $n, m \ge 0$ there exists such $s_1, s_2, s_3, s_4 \in \mathbb{Z}$ that,
	\be\begin{split}
		\vartheta(\mb{p}, m) &= \prod_{i=1}^4 T_i^{s_i} (\vartheta(\mb{c}, n)) \\
		\mbox{and therfore, } \quad \vartheta(\mb{p}, m+1) &= \prod_{i=1}^4 T_i^{s_i} (\vartheta(\mb{c}, n+1))
	\end{split}\ee
	Then according to (\ref{f_psi}) we can write:
	\be\begin{split}
		f(\vartheta(\mb{c}, n), \vartheta(\mb{p}, m)) &= \psi\lt( \mb{c}, n, \{s_i\}_{i=1}^4 \rt) \\
		\mbox{and,} \quad f(\vartheta(\mb{c}, n+1), \vartheta(\mb{p}, m+1)) &= \psi\lt( \mb{c}, n+1, \{s_i\}_{i=1}^4 \rt)
	\end{split}\ee
	Comparing (\ref{master_rec_L}) with (\ref{master_rec_1}) we see that for fixed $\mb{c}$ and $\{s_i\}$, $\psi\lt( \mb{c}, n, \{s_i\}_{i=1}^4 \rt)$ is at most a linear function of $n$ and therefore,
	\be\begin{split}
		\psi\lt( \mb{c}, n+1, \{s_i\}_{i=1}^4 \rt) &\sim \psi\lt( \mb{c}, n, \{s_i\}_{i=1}^4 \rt) \quad [\mbox{as } n \to \infty] \\
		\Rightarrow f(\vartheta(\mb{c}, n+1), \vartheta(\mb{p}, m+1)) &\sim f(\vartheta(\mb{c}, n), \vartheta(\mb{p}, m)) \quad [\mbox{as } n \to \infty]
	\end{split}\ee
	Using the definition of $\mathcal{T}$ in (\ref{transition_Nd}) the property of $f$ in the above equation can be transferred to $\mathcal{T}$ and we get,
	\be
		\mathcal{T}(\vartheta(\mb{c}, n+1), \vartheta(\mb{p}, m+1)) \sim \mathcal{T}(\vartheta(\mb{c}, n), \vartheta(\mb{p}, m)) \quad [\mbox{as } n \to \infty] \label{T_translation}
	\ee
	In particular, this remains valid for $\mathcal{T}'$ as well. Now, using (\ref{sim_1}), (\ref{sim_2}), (\ref{T''}) and (\ref{T_translation}) all together we get for any $\mb{c}, \mb{q} \in B_0$, $\mb{p} \in \widetilde B_{34} \subset B_0$, and $n \ge 0$:
	\bea
		&& \sum_{m=0}^\infty \mathcal{T}'(\vartheta(\mb{c}, n), \vartheta(\mb{p}, m)) \mathcal{T}''(\vartheta(\mb{p}, m), \mb{q}) \nn\\
		&\sim& \sum_{m=0}^\infty \mathcal{T}'(\vartheta(\mb{c}, n+1), \vartheta(\mb{p}, m+1)) \mathcal{T}''(\vartheta(\mb{p}, m+1), \mb{q}) \rho^{-1} \quad [\mbox{as } n \to \infty] \nn\\
		&=& \sum_{m=1}^\infty \mathcal{T}'(\vartheta(\mb{c}, n), \vartheta(\mb{p}, m)) \mathcal{T}''(\vartheta(\mb{p}, m), \mb{q}) \rho^{-1} \label{temp_1}
	\eea
	Equation (\ref{L_to_B1_B2_4}) and (\ref{temp_1}) enable us to evaluate the limit in (\ref{conv_cond}) as follows:
	\be
		\lim_{n \to \infty} \frac{\s(\vartheta(\mb{c},n+1))}{\s(\vartheta(\mb{c},n))} = \lim_{n \to \infty} \;
		\frac{\sum_{\substack{ \mb{q} \in B_0 \\ \mb{p} \in \widetilde B_{34} }} \mathcal{T}'(\vartheta(\mb{c}, n), \vartheta(\mb{p}, 0)) \mathcal{T}''(\vartheta(\mb{p}, 0), \mb{q}) \s(\mb{q})}
		{\sum_{\substack{ \mb{q} \in B_0 \\ \mb{p} \in \widetilde B_{34} }} \sum_{m=0}^\infty \mathcal{T}'(\vartheta(\mb{c}, n), \vartheta(\mb{p}, m)) \mathcal{T}''(\vartheta(\mb{p}, m), \mb{q}) \s(\mb{q})} + \rho = \rho
	\ee
	The last equality is achieved because the sum over $m$ in the denominator of the middle term is divergent. Because the highest value of $m$ for which we have nonzero summand depends linearly on $n$ and at the $n \to \infty$ limit the sum is truly an infinite sum with undamped summands. Therefore we reach at the following condition for divergence of $V_{\mb{c}}$:
	\bea
		\lim_{n \to \infty} \frac{\s(\vartheta(\mb{c},n+1))}{\s(\vartheta(\mb{c},n))} \eta &=& 1 \nn\\
		\mbox{or,} \qquad \rho h \frac{\ln u}{u} &=& 1 \label{the_singularity}
	\eea

\subsection{Comments on the singularity}
	The result of the previous section shows that the singularity of the effective potential of MSED is not the usual Landau pole $u = 1 - \frac{\ln(\phi/\mu)}{w} = 0$ as suggested by perturbative expressions of the potential in (\ref{e72}), (\ref{e74}). Written in a more explicit form, (\ref{the_singularity}) becomes:
	\be
		\frac{-b_{3,0}^g}{b_{2,0}^g} \frac{1}{b_{2,0}^g w} \frac{\ln \lt(1 - \frac{\ln(\phi/\mu)}{w}\rt)}{1 - \frac{\ln(\phi/\mu)}{w}} = 1 \label{the_singularity_explicit}
	\ee
	Where we have used (\ref{k2inbeta}) and the fact that $b_{\_,0}^\alpha = 0$. To solve eq. (\ref{the_singularity_explicit}) for $\phi$ we need to know, of all the terms of the beta functions, only $b_{3,0}^g$ and $b_{2,0}^g$ which come from just the 1-loop and 2-loop calculations. We also note that this result is fully consistent with the previous results obtained for the scalar $\phi^4$ theory \cite{summing, unique} to which it reduces in the limit $\alpha \to 0$.

\section{POSSIBLE EXTENSION TO MULTIPLE COUPLINGS}
	In this section we look at possible extensions of our result for theories with more than one coupling constants. We make
	some assumptions along the way. One is that, the quartic interaction of the scalar field is the common one.
	
	Now suppose the coupling constants in some theory are denoted $\lm_1, \cdots, \lm_n$ and $\lm_n$ is the quartic coupling. Also
	suppose, the 1-loop beta function for $\lm_i$, denoted as $\beta_2^{\lm_i}$, depends only on $\lm_j$ such that $1 \leq j \leq i$.
	We note that this is true for the Standard Model where $\lm_n$ corresponds to the quartic Higgs self interaction. We define the characteristic
	functions $\tilde \lm_i(t)$ as solutions to the differential equations:
	\be
		\frac{\dd \tilde \lm_i(t)}{\dd t} = \beta_2^{\lm_i} \lt(\tilde\lm_1(t), \cdots, \tilde\lm_i(t)\rt) \label{gbeta}
	\ee
	If $\tilde\lm_n(t)$ has a simple pole at $t = w$ then, using (\ref{gbeta}) it is easy to deduce that:
	\be
		\lt.\frac{\dd^m \tilde \lm_i}{\dd t^m}\rt|_{t=w} := C_{m+1}^i \in \mathbb{R}_{m+1}\lt[ \tilde\lm_1(w), \cdots, \tilde\lm_i(w) \rt];\quad [i < n] \label{lmidrv}
	\ee
	where $K_m[x_1, \cdots, x_k]$ denotes the vector space of homogeneous polynomials of degree $m$ in the variables
	$x_1, \cdots, x_k$ over the field $K$. Using the assumption that $w=0$ is a simple pole of $\tilde \lm_n(t)$ we also find that,
	\be
		\lt.\frac{\dd^m}{\dd t^m} \lt( (t-w) \tilde \lm_n \rt)\rt|_{t=w} := \lim_{t' \to w} \lt[\lt.\frac{\dd^m}{\dd t^m} \lt( (t-w) \tilde \lm_n \rt)\rt|_{t=t'}\rt] := C_m^n \in \mathbb{R}_m\lt[ \tilde\lm_1(w), \cdots, \tilde\lm_{n-1}(w) \rt] \label{lmndrv}
	\ee
	Equations (\ref{lmidrv}) and (\ref{lmndrv}) can be used to find the Laurent expansion for $\lm_i(t)$ around $t = w$:
	\bea
		\lm_i(t) &=& \sum_{m=0}^\infty \frac{C_{m+1}^i}{m!} (t-w)^m\; ;\quad [i < n] \label{taylori} \\
		\lm_n(t) &=& \sum_{m=-1}^\infty \frac{C_{m+1}^n}{(m+1)!} (t-w)^m \label{taylorn}
	\eea
	Equations (\ref{taylori}) and (\ref{taylorn}) are direct generalizations of equations (\ref{e65}) and (\ref{e66}) and these
	forms are essential for further calculations towards the singularity structure. Another important assumption is that we can
	write the effective potential of the theory in the following form:
	\be
		V = \sum_{i=1}^\infty \sum_{j=0}^{i-1} P_i^j (\lm_1, \cdots, \lm_n) L^j \phi^4;\quad [L = \ln\frac{\phi^2}{\mu^2}]
	\ee
	where $P_i^j (\lm_1, \cdots, \lm_n) \in \mathbb{R}_i[\lm_1, \cdots, \lm_{n-1}]$ and $\phi$ is the scalar field with the
	quartic interaction. The method of characteristics readily generalizes for arbitrary couplings. We modify the definition
	in (\ref{e95}) as follows:
	\be
		\,[m]_k^t := \sum_{i=k}^\infty c_{i+m} \lt(\tilde\lm_1(w), \cdots, \tilde\lm_n(w)\rt) (t-w)^i
	\ee
	where $c_{i+m} \lt(\tilde\lm_1(w), \cdots, \tilde\lm_n(w)\rt) \in \mathbb{R}_{i+m} \lt[\tilde\lm_1(w), \cdots, \tilde\lm_n(w)\rt]$. Now, if with this modified notation equations (\ref{e71}) and (\ref{e53}) remain unchanged then we can proceed as follows. As in (\ref{e56}) we make the change of variables $$(\lm_1, \cdots, \lm_n) \rightarrow (\chi_1, \cdots, \chi_{n-1}, h) := \lt(\tilde \lm_1(w), \cdots, \tilde \lm_{n-1}(w), 1/w \rt).$$
	Change comes in (\ref{simplifyprod}) where instead of just $\chi^N$ we would now have a homogeneous polynomial of degree $N$
	in the variables $\chi_1, \cdots, \chi_{n-1}$. Therefore we would have as the analog of $S_{N,P}$ from equation (\ref{e72})
	something denoted as $S_{\mathbf{N}, P}$ where $\mathbf{N}$ is a $(n-1)$ dimension vector with positive integer components,
	i.e., $\mathbf{N} = (N^1, \cdots, N^{n-1}) \in \mathbb{Z}_{\ge 0}^{n-1}$. But the form of $S_{N,P}$ as can be seen on the
	right hand side of equation (\ref{e72}) and that of the generalized $S_{\mathbf{N}, P}$ will essentially be the same with the
	$N$ in the r.h.s. of (\ref{e72}) replaced by $||\mathbf{N}|| := \sum_i N^i$. Since this particular form of $S_{N,P}$ was what enabled
	the analysis in section \ref{secREC}, we would have a recurrence relation to determine the potential with the new $k$'s defined by the following equation:
	\be
		\sum_{i=1}^n \beta^{\lm_i} \frac{\p h}{\p \lm_i} = \sum_{n=2}^{\infty}\, \sum_{\substack{\;\mathbf{p} \in \mathbb{Z}_{\ge 0}^n \\ ||\mathbf{p}|| = n}} k_\mathbf{p}^n\, h^{p^1} \chi_1^{p^2} \cdots \chi_{n-1}^{p^n}
	\ee
	The constructions of Section \ref{sec_singularity_analysis} are readily generalizable as well. We can partition the lattice space $L$ by introducing an equivalence relation and partition $V$ as in (\ref{V_part}) by introducing $V_{\mb{c}}$ where $\mb{c}$ will belong to an equivalence class like $B_0$. And the independence of the singularity from the constants whose powers remain bounded by the choice of $\mb{c}$ (denoted as elements of $\Omega$ in (\ref{Omega})) and does not depend on $n$ (as in \ref{V_part})) decouples parts of $V$ arising from different powers of $\chi_i$ and we reach at a much simpler recurrence relation than the original one, as was done in (\ref{master_rec_1}). Asymptotic analysis makes further simplifications and the condition for divergence remains similar in form to (\ref{the_singularity}). Only changes are that $b_{3,0}^g$ and $b_{2,0}^g$ will be replaced by $b_{3,0,\cdots,0}^g$ and $b_{2,0,\cdots,0}^g$ respectively which still come from the 1-loop and 2-loop beta functions only.

\section{AN ALTERNATE SUMMATION}
	In this section we follow \cite{improv}, showing that in the CW scheme the RG improved effective potential becomes
	independent of $\phi$. We write $V$ in the form,
	\be V = Y(\alpha, g, l)\phi^4;\qquad \lt(l = \frac{L}{2} = \mathrm{ln}\frac{\phi}{\mu}\rt) \label{e41}\ee
	and regroup the terms of (\ref{e40}) in a way that,
	\be Y(\alpha, g, l) = \sum_{n=0}^{\infty} A_n(\alpha, g) l^n. \label{e42}\ee
	Substitution of (\ref{e41}) and (\ref{e42}) in (\ref{e6}) gives the recurrence relation:
	\be (n+1)A_{n+1} = \lt( \hat\beta^g \frac{\p}{\p g} + \hat\beta^{\alpha} \frac{\p}{\p\alpha} + 4\hat\gamma \rt) A_n \ee
	where, \be \hat\beta^g = \frac{\beta^g}{1-\gamma};\quad \hat\beta^{\alpha} = \frac{\beta^{\alpha}}{1-\gamma};\quad \hat\gamma = \frac{\gamma}{1-\gamma}. \nn\ee
	Now define $\bva(t)$, $\bvg(t)$ and $\bvA(\bva(t), \bvg(t), t)$ such that,
	\bea
		\frac{\dd\bva(t)}{\dd t} &=& \hat\beta^{\alpha}(\bva, \bvg)\quad \mbox{with } \bva(0) = \alpha \label{e45}\\
		\frac{\dd\bvg(t)}{\dd t} &=& \hat\beta^{g}(\bva, \bvg)\quad \mbox{with } \bvg(0) = g \label{e46}\\
		\mbox{and, } \bvA(\bva, \bvg, t) &=& A_n(\bva, \bvg)\, \mathrm{exp}\lt[4\int_0^t \hat\gamma(\bva(\tau), \bvg(\tau))\, \dd\tau\rt] \nn
	\eea
	Then,
	\bea
		\frac{\dd\bvA_n}{\dd t} &=& \lt( \hat\beta^g (\bva, \bvg) \frac{\p}{\p\bvg} + \hat\beta^{\alpha} (\bva, \bvg) \frac{\p}{\p\bva} + 4\hat\gamma (\bva, \bvg) \rt) \bvA_n = (n+1) \bvA_{n+1} \label{e44}\\
		\Rightarrow \bvA_{n+1} &=& \frac{1}{n+1} \frac{\dd\bvA_n}{\dd t} = \frac{1}{(n+1)!} \frac{\dd^{n+1}\bvA_0}{\dd^{n+1} t} \label{e43}
	\eea
	Define, \be \breve Y(\bva(t), \bvg(t), l, t) = \sum_{n=0}^{\infty} \bvA_n(\bva, \bvg, t) l^n \nn\ee
	so that, \be \breve Y(\bva(0), \bvg(0), l, 0) = Y(\alpha, g, l) \nn\ee
	Now,
	\bea
		\breve Y(\bva(t), \bvg(t), l, t) &=& \sum_{n=0}^{\infty} \bvA_n(\bva, \bvg, t) l^n \nn\\
		&=& \sum_{n=0}^{\infty} \frac{l^n}{n!} \frac{\dd^n\bvA_0}{\dd^n t};\qquad \mbox{using }(\ref{e43}) \nn\\
		&=& \bvA_0(\bva(t+l), \bvg(t+l), t+l) \nn\\
		&=& A_0(\bva(t+l), \bvg(t+l))\, \mathrm{exp}\lt[4\int_0^{t+l} \hat\gamma(\bva(\tau), \bvg(\tau))\, \dd\tau\rt] \label{e48}
	\eea
	
	If now the renormalization scale $\mu$ is taken to be equal to the vacuum expectation value of the scalar field $\phi$, denoted by $v$, i.e.,
	\be \lt.\frac{\dd V}{\dd\phi}\rt|_{\phi = \mu} = 0 \ee
	then from (\ref{e41}) and (\ref{e42}) and noting that $l=0$ at $\phi=\mu$ we get,
	\be (A_1 + 4A_0)\, v^3 = 0 \nn\ee
	This coupled with (\ref{e44}) gives (for $v \neq 0$),
	\be \lt(\hat\beta^g \frac{\p}{\p g} + \hat\beta^{\alpha} \frac{\p}{\p \alpha} + 4(1+\hat\gamma) \rt)A_0 = 0 \label{e47}\ee
	Using (\ref{e45}) and (\ref{e46}) the parametric solution of (\ref{e47}) is,
	\be A_0(\bva(t), \bvg(t)) = A_0(\bva(0), \bvg(0))\, \mathrm{exp} \lt[-4\int_0^t (1+\hat\gamma(\bva(\tau), \bvg(\tau))) \dd\tau \rt] \label{e49}\ee
	Substituting (\ref{e49}) in (\ref{e48}) we get,
	\bea
		\breve Y(\bva(t), \bvg(t), l, t) &=& A_0(\bva(0), \bvg(0))\, \mathrm{exp} \lt[-4\int_0^{t+l} (1+\hat\gamma(\bva(\tau), \bvg(\tau))) \dd\tau \rt] \mathrm{exp}\lt[4\int_0^{t+l} \hat\gamma(\bva(\tau), \bvg(\tau))\, \dd\tau\rt] \nn\\
		&=& A_0(\alpha, g)\, \mathrm{exp}\lt[-\int_0^{t+l}4\, \dd\tau\rt] \nn\\
		&=& A_0(\alpha, g)\, e^{-4(t+l)} \nn\\
		\Rightarrow Y(\alpha, g, l) &=& \breve Y(\bva(0), \bvg(0), l, 0) = A_0(\alpha, g)\, e^{-4l} \nn\\
		&=& A_0(\alpha, g)\,\frac{\mu^4}{\phi^4} \nn\\
		\Rightarrow V &=& A_0(\alpha, g)\,\mu^4\qquad \mbox{using } (\ref{e41}) \label{e50}
	\eea
	From (\ref{e50}) it is evident that $V$ is independent of $\phi$.

\section{DISCUSSION}
	Re-summation of the leading-logarithm contributions  $V_{m}$  to the effective potential in such a way that portions of $V$ beyond some particular value of $m$ being included in that sum is always possible and can be obtained as exact closed form expressions \cite{unique, summing}.  When we go beyond some particular order of estimation or some specific value of $m$ by that re-summation, the  singularity in
	$V$, which shows itself as a singularity in the individual leading order summations, is shifted away from the "Landau Singularity" revealing  a peculiar singularity structure \cite{unique}. This interesting feature might have a role to play in the standard model and the possibility is an open question which may be worth pursuing \cite{higgs}. In this paper, a non-trivial extension of this re-summation to a theory with multiple couplings is accomplished, which can further refine the estimates of Higgs mass\cite{higgs} in the conformal limit of the standard model,
	where, up to $m=4$ (up to five-loop order) have been used. In fact this result is improved up to nine-loop order in \cite{tom} using Pade' approximations and an averaging method resulting in an upper bound on the Higgs mass of 141 GeV. More rigorous estimates
will need to follow the re-summation of the effective potential (which shifts the "Landau" singularity) that we dealt with or summing
$V_{m}$ beyond $m=4$ using the techniques of \cite{higgs}. But in both cases, the knowledge of the exact three-loop RG functions and beyond for the standard model is necessary which have not been calculated yet.These estimates (alongside enhanced Higgs-Higgs scattering processes)\cite{higgs,tom} are very interesting as they are consistent with the recent ATLAS and CMS Collaborations observations of the 125 GeV Higgs mass and provide signals to distinguish conventional and radiative electro-weak summetry breaking\cite{tom}. In other words, the Coleman-Weinberg approach really seems to be viable and hence its pressing to shed light on the formal structure of conformally invariant models. The most important thing is the result that the singularity structure of the complete effective potential of MSED or standard model in the conformal limit is completely different from their perturbative approximations, consistent with \cite{summing}. In fact, the singularity structure of the effective potential is not deducible from any order of perturbative treatment, however large. Most interestingly, it depends only on the 1-loop and 2-loop beta functions of the quartic scalar coupling. The apparent
	peculiarity of the singularity structure or the flatness of the total effective potential might indicate shortcomings of the applicability of perturbation theory in the context or something completely unknown, because the effective potential that we are talking about is the one relevant for analyzing spontaneous symmetry breaking. We have also outlined a rigorous framework  for the application of our results  to theories with arbitrary number of couplings, which in particular includes the standard model .
	
	%\clearpage
\section*{Acknowledgements}

%This work was largely inspired by the late Victor Elias.  Roger Macleod had a useful suggestion.  There was helpful correspondence with C.~Ford and S.~Martin. NSERC (Natural Science \& Engineering Research Council of Canada)  provided funding for RBM and TGS.
	TH would like to thank Dr. D.G.C McKeon and Dr.T.G.Steele for their helpful suggestions.
		
%\appendix
%\appendixpage
\section*{Appendix 1: Properties of $[m]_k^t$}\label{app1}
%\section{Properties of $[m]_k^t$}\label{app1}
	We illustrate different operations involving the entities $[m]_k^t$ introduced in (\ref{e95}):
	\begin{itemize}
		\item \textbf{Multiplication:} \bea
			[m]_k^t[n]_l^t &=& \lt(\sum_{i=k}^{\infty} c_i^1\,\ta(w)^{i+m} (t-w)^i\rt) \lt(\sum_{j=l}^{\infty} c_j^2\,\ta(w)^{j+n} (t-w)^j\rt) \nn\\
			&=& \sum_{p=k+l}^{\infty} c_p^3\,\ta(w)^{p+m+n} (t-w)^p = [m+n]_{k+l}^t \nn
		\eea
		\item \textbf{Addition:}\quad $[m]_k^t + [m]_l^t = [m]_{\mathrm{min}(k,l)}^t$
		\item \textbf{Multiplication by a number:}\quad $c[m]_k^t = [m]_k^t$;\quad $c \in \mathbb{R}$
		\item \textbf{Inverse:}\quad $([m]_k^t)^{-1} = \ta(w)^{-m-k}(t-w)^{-k}\,([0]_0^t)^{-1} = \ta(w)^{-m-k}(t-w)^{-k}\,[0]_0^t = [-m]_{-k}^t$
		\item \textbf{Integration:}\quad Integration of the series $[m]_k^t$ with respect to $t$ results in another series, the
			form of which depends on the parameter $k$. The integrals are as follows:\\
		
			$k>-1$: \bea \int [m]_k^{\tau}\, \dd\tau &=& \sum_{i=k}^{\infty} c_i\,\ta(w)^{i+m} \int (\tau-w)^i \dd\tau \nn\\
				&=& \sum_{i=k}^{\infty} c'_i\,\ta(w)^{i+m} (\tau-w)^{i+1} \dd\tau;\quad c'_i = \frac{c_i}{i+1} \nn\\
				&=& \sum_{i=k+1}^{\infty} c''_i\,\ta(w)^{i+m-1} (\tau-w)^i \dd\tau;\quad c''_i = c'_{i-1} \nn\\
				&=& [m-1]_{k+1}^{\tau} \nn
			\eea
			$k=-1$: \bea \int [m]_{-1}^{\tau}\, \dd\tau &=& c_{-1}\,\ta(w)^{m-1} \int (\tau-w)^{-1} \dd\tau + \int [m]_0^{\tau}\, \dd\tau \nn\\
					&=& c_{-1}\,\ta(w)^{m-1} \mathrm{ln} (\tau-w) + [m-1]_1^{\tau} \nn
				\eea
			$k<-1$: \bea \int [m]_k^{\tau}\, \dd\tau &=& c_{-1}\,\ta(w)^{m-1} \int (\tau-w)^{-1} \dd\tau + \int [m]_k^{\tau}\, \dd\tau \nn\\
					&=& c_{-1}\,\ta(w)^{m-1} \mathrm{ln} (\tau-w) + [m-1]_{k+1}^{\tau} \nn
				\eea
	\end{itemize}
\section*{Appendix 2: Derivatives of $y_2$}
%\section{Derivatives of $y_2$}
	We rewrite here the PDE from eq. (\ref{e21}):
	\bea
		&& \beta_2^{x_1} \frac{\p y_2}{\p\tx_1} + \beta_2^{x_2} \frac{\p y_2}{\p\tx_2} = \frac{y_2^2}{wg} \nn\\
		\Rightarrow && \frac{\tx_1^2}{24\pi^2} \frac{\p y_2}{\p\tx_1} + \frac{5 \tx_2^2 - 12 \tx_2\tx_1 + 24\tx_1^2}{16\pi^2} \frac{\p y_2}{\p\tx_2} = \frac{y_2^2}{wg} \nn
	\eea
	and $y_2$ must satisfy $y_2(x_1, x_2) = \frac{16\pi^2}{5w}$.
	In an attempt to solve the PDE, we make the change of variables $(\tx_1, \tx_2) \rightarrow (\tx_1, x_3)$; where,
	$$x_3(\tx_1, \tx_2) = \frac{\tx_1}{\tx_2}.$$
	Now, we write the PDE in terms of $\tx_1$ and $x_3$:
	\be
		\frac{\tx_1^2}{24\pi^2} \frac{\p y_2}{\p\tx_1} - \frac{\tx_1(5 - 12 x_3 + 24x_3^2)}{16\pi^2} \frac{\p y_2}{\p x_3} = \frac{y_2^2}{wg} \label{e91}
	\ee
	The form of solution for this PDE is, $y_2(\tx_1, x_3) = \tx_1Y(x_3)$. Substituting this form in (\ref{e91}) we get,
	\be
		\frac{Y}{24\pi^2} - \frac{5 - 12 x_3 + 24x_3^2}{16\pi^2} \frac{\dd Y}{\dd x_3} = \frac{Y^2}{wg} \label{e92}
	\ee
	This is an ODE for $Y$ and a solution with the desired boundary condition exists. We employ the Frobenius method to find the
	solution. We seek a power series solution of the form:
	\be Y(x_3) = \sum_{i=s}^\infty A_ix_3^i;\quad [A_s \neq 0] \nn\ee
	If we substitute this form in (\ref{e92}), then the lowest order term in the LHS is $-\frac{5}{16\pi^2}sA_sx_3^{s-1} \neq 0$
	and the lowest order term in the RHS is $\frac{1}{wg} A_s^2x_3^{2s} \neq 0$. These two terms must be equal. Therefore,
	$$2s = s-1 \Rightarrow s = -1.$$ Hence, $y_2$ has the form,
	\be
		y_2 = \tx_1Y(x_3) \;=\; \tx_1 \sum_{i=-1}^\infty A_ix_3^i = \tx_1 \sum_{i=-1}^\infty A_i \lt( \frac{\tx_1}{\tx_2} \rt)^i = \sum_{i=-1}^\infty A_i \frac{\tx_1^{i+1}}{\tx_2^i} \nn
	\ee
	We shall not try to find the coefficients $A_i$ here, rather we shall look at the derivatives of $y_2$ w.r.t. its arguments.
	We notice,
	\bea
		\frac{\p y_2}{\p \tx_1} &=& \sum_{i=-1}^\infty (i+1)A_i \lt( \frac{\tx_1}{\tx_2} \rt)^i = \sum_{i=-1}^\infty (i+1)A_i \lt( \frac{[1]_0^t}{[1]_{-1}^t} \rt)^i = \sum_{i=-1}^\infty (i+1)A_i ([0]_1^t)^i \nn\\
		&=& \sum_{i=-1}^\infty (i+1)A_i [0]_i^t \; =\; \sum_{i=0}^\infty (i+1)A_i [0]_i^t \; =\; \sum_{i=0}^\infty [0]_i^t = [0]_0^t \label{e93}
	\eea
	Similarly we also find,
	\be
		\frac{\p y_2}{\p \tx_2} = [0]_0^t \label{e94}
	\ee
	%The solution is:
	%\be y_2(\tx_1, \tx_2) = \tx_1\lt[\frac{24\pi^2}{gw} + \mathrm{exp}\lt\{ \frac{2}{\sqrt{719}} \lt(\tan^{-1}\frac{15(\tx_2/\tx_1)-19}{\sqrt{719}} - \tan^{-1}\frac{15(x_2/x_1)-19}{\sqrt{719}} \rt) \rt\} \lt( \frac{5x_1w}{16\pi^2} - \frac{24\pi^2}{wg} \rt)\rt]^{-1} \nn\ee

\section*{Appendix 3: Relating $k$ to the beta functions} \label{k_to_beta}
%\section{Relating $k$ to the beta functions} \label{k_to_beta}
	Here we find expressions of $k_{\_,\_}^1$, $k_{\_,\_}^2$ and $k_{\_,\_}^3$ defined by eq. (\ref{e62}) in terms of $b_{\_,\_}^g$, $b_{\_,\_}^\alpha$ and $\tau_{\_,\_}$ defined by (\ref{e7}). From (\ref{e73}),
	\be\begin{split}
		\alpha &= \chi\,\bar f_1\lt(\frac{\chi}{h}\rt) = \chi \sum_{i=0}^\infty A_i\, \chi^i h^{-i} \\
		g &= h\,\bar f_2\lt(\frac{\chi}{h}\rt) = h \sum_{i=0}^\infty B_i\, \chi^i h^{-i}
	\end{split}\label{aginch}\ee
	Now, beginning from (\ref{e7}a):
	\be
		\beta^g = \sum_{n=2}^{\infty} \sum_{r=0}^n b_{n-r,r}^g g^{n-r} \alpha^r = \sum_{n=2}^{\infty} \sum_{k=0}^\infty \sum_{r=0}^{\mathrm{min}(n,k)} b_{n-r,r}^g E_{k-r}^g\, \chi^k h^{n-k};\quad [k = r+i] \nn
	\ee
	Comparing the last line with (\ref{e85}a) we find,
	\be \bar b_{n,k}^g = \sum_{r=0}^{\mathrm{min}(n,k)} b_{n-r,r}^g E_{k-r}^g \label{e88}\ee
	Similarly, beginning from (\ref{e7}b) we shall find,
	\be
		\beta^\alpha = \sum_{n=2}^\infty \sum_{k=0}^\infty \bar b_{n,k}^\alpha \chi^k h^{n-k} \qquad \mbox{with,} \qquad \bar b_{n,k}^\alpha = \sum_{r=0}^{\mathrm{min}(n,k)} b_{n-r,r}^\alpha E_{k-r}^\alpha \label{e89}
	\ee
	From (\ref{e56}),
	\begin{alignat}{1}
		h = \sum_{i=0}^\infty C_i^h \alpha^i g^{-i+1} \qquad &\mbox{and,} \qquad \chi = \sum_{i=0}^\infty C_i^\chi \alpha^{i+1} g^{-i}\nn\\
		\Rightarrow \frac{\p h}{\p g} = \sum_{i=0}^\infty C_i^{h,g} \chi^i h^{-i} \qquad &\mbox{and,} \qquad \frac{\p \chi}{\p g} = \sum_{i=1}^\infty C_i^{\chi,g} \chi^{i+1} h^{-(i+1)} \nn\\
		\mbox{Also,} \quad \frac{\p h}{\p \alpha} = \sum_{i=1}^\infty C_i^{h,\alpha} \chi^{i-1} h^{-(i-1)} \qquad &\mbox{and,} \qquad \frac{\p \chi}{\p \alpha} = \sum_{i=0}^\infty C_i^{\chi,\alpha} \chi^i h^{-i} \nn
	\end{alignat}
	Now,
	\be
		\beta^g \frac{\p h}{\p g} = \sum_{n=2}^\infty \sum_{k=0}^\infty \sum_{i=0}^\infty \bar b_{n,k}^g C_i^{h,g} \chi^{i+k} h^{n-i-k} = \sum_{n=2}^\infty \sum_{j=0}^\infty \lt( \sum_{k=0}^j \bar b_{n,k}^g C_{j-k}^{h,g} \rt) \chi^j h^{n-j} \label{bgphg}
	\ee
	Similarly we find,
	\bea
		\beta^\alpha \frac{\p h}{\p \alpha} &=& \sum_{n=2}^\infty \sum_{j=0}^\infty \lt( \sum_{k=0}^j \bar b_{n,k}^\alpha C_{j-k+1}^{h, \alpha} \rt) \chi^j h^{n-j} \label{bapha} \\
		\beta^g \frac{\p \chi}{\p g} &=& \sum_{n=2}^\infty \sum_{j=1}^\infty \lt( \sum_{k=0}^{j-1} \bar b_{n,k}^g C_{j-k}^{\chi, g} \rt) \chi^{j+1} h^{n-j-1} \label{bgpcg} \\
		\beta^\alpha \frac{\p \chi}{\p \alpha} &=& \sum_{n=2}^\infty \sum_{j=0}^\infty \lt( \sum_{k=0}^j \bar b_{n,k}^\alpha C_{j-k}^{\chi, \alpha} \rt) \chi^j h^{n-j} \label{bacpa}
	\eea
	Adding (\ref{bgpcg}) with (\ref{bacpa}) and comparing with (\ref{e62}a) we get,
	\be
		k_{n-j,j}^1 = \sum_{k=0}^{j-2} \bar b_{n,k}^g C_{j-1-k}^{\chi,g} + \sum_{k=0}^j \bar b_{n,k}^\alpha C_{j-k}^{\chi, \alpha}
		= \sum_{k=0}^{j-2} \sum_{r=0}^{\mathrm{min}(n,k)} b_{n-r,r}^g E_{k-r}^g C_{j-1-k}^{\chi,g} + \sum_{k=0}^j \sum_{r=0}^{\mathrm{min}(n,k)} b_{n-r,r}^\alpha E_{k-r}^\alpha C_{j-k}^{\chi,\alpha} \label{k1inbeta}
	\ee
	To get the last line we used (\ref{e88}) and (\ref{e89}). Similarly, adding (\ref{bgphg}) with (\ref{bapha}) and comparing with (\ref{e62}b) we get,
	\be
		k_{n-j,j}^2 = \sum_{k=0}^j \lt( \bar b_{n,k}^g C_{j-k}^{h,g} + \bar b_{n,k}^\alpha C_{j-k+1}^{h, \alpha} \rt)
		= \sum_{k=0}^j \sum_{r=0}^{\mathrm{min}(n,k)} \lt( b_{n-r,r}^g E_{k-r}^g C_{j-k}^{h,g} + b_{n-r,r}^\alpha E_{k-r}^\alpha C_{j-k+1}^{h,\alpha} \rt) \label{k2inbeta}
	\ee
	Substitution of (\ref{aginch}) in (\ref{e7}c) leads to the expression for $k_{n-j, j}^3$:
	\be
		k_{n-j, j}^3 = \sum_{k=0}^{\mathrm{min}(j,n)} \tau_{n-k,k} D_k \label{k3ingamma}
	\ee
	where $D$ can be determined from $A$ and $B$ of (\ref{aginch}). (\ref{k1inbeta}), (\ref{k2inbeta}) and (\ref{k3ingamma}) shows that $k^i_{j,k}$ depends only on such $b_{m,n}^1$, $b_{m,n}^2$ and $\tau_{m,n}$ that $m+n = j+k$. But we recall from the definition of these terms (eq. (\ref{e7})) that the sum of the subscripts denote the order of loop of the feynman diagrams that goes into their calculation. This shows that $k^i_{j,k}$ depends on the $(j+k)$-loop beta and gamma functions.

\end{document}